\documentclass[global,twocolumn]{svjour}
\journalname{}
\usepackage{latexsym}
\usepackage[english]{babel}
\usepackage{graphicx} 
\usepackage{subfig} 
\begin{document}
\title{Linear Paul trap design for an optical clock with Coulomb crystals}
\author{N.~Herschbach\inst{1}, K.~Pyka\inst{1}, J.~Keller\inst{1} \and T.~E.~Mehlst\"aubler\inst{1}}
\institute{Physikalisch-Technische Bundesanstalt, Bundesallee 100, 38116 Braunschweig, Germany}
\date{Received: date / Revised version: date}
\maketitle
\begin{abstract}
We report on the design of a segmented linear Paul trap for
optical clock applications using trapped ion Coulomb crystals. For
an optical clock with an improved short-term stability and a
fractional frequency uncertainty of $10^{-18}$, we propose
$^{115}$In$^+$ ions sympathetically cooled by $^{172}$Yb$^+$. We
discuss the systematic frequency shifts of such a frequency
standard. In particular, we elaborate on high precision calculations of
the electric radiofrequency field of the ion trap using the finite
element method. These calculations are used to find a scalable
design with minimized excess micromotion of the ions at a level at
which the corresponding second-order Doppler shift contributes
less than $10^{-18}$ to the relative uncertainty of the frequency
standard.
\end{abstract}
\section{Introduction}\label{intro}
Since the first demonstration of a single trapped ion in 1980,
experiments with trapped single particles have led to some of the
finest spectroscopic measurements in
physics~\cite{Neuhauser,Werth,Gabrielse} and radiofrequency (rf)
Paul traps have become outstanding working tools with unsurpassed
accuracy in frequency
metrology~\cite{Bergquist_Rosenband_Science,Chou}.

The location of the single ion and thus, electro-magnetic fields
sensed by it, can be controlled and measured at the nm scale,
leading to a superior precision of optical ion clocks with a
fractional systematic frequency uncertainty now reaching down into
the $10^{-18}$ regime~\cite{Chou}. Still, the low signal-to-noise
ratio of the single ion interrogation limits the obtainable
short-term stability and puts up high demands for the stability of
the optical clock laser. One approach to improve the short-term
stability of ion clocks is to investigate narrower atomic
transitions with mHz linewidth and thus, higher quality factor.
Here, for Fourier limited spectroscopy a short-term linewidth of
the clock laser at the mHz level is required. Due to the limited
quantum information obtained in single ion spectroscopy, long
integration times are necessary to lock the laser onto the atomic
signal. For example, resolving and locking a laser to a mHz wide
clock transition by standard quantum jump spectroscopy, would
require a clock laser frequency stability in the low $10^{-17}$
range over several minutes, in order to reach the quantum
projection noise limit of the single ion~\cite{Itano}. A major
effort is made in the metrology community to push the stability
limits of optical reference systems~\cite{Meiser,Millo,Legero} and
faster, phase sensitive detections schemes could be thought of.
Still, today's state-of-the-art clock laser stability and
detection limit the achieved short term stability such, that
integration times of tens of days to weeks would be necessary to
reach a frequency resolution of $1 \times 10^{-18}$ with an
optical ion clock. This high frequency resolution is required for
improved fundamental tests of physics and applications in
geodesy~\cite{Bjerhammer,Pavlis}. As for white laser noise, the
stability of the frequency measurement averages down with the
square root of both atom number N and integration time
$\tau$~\cite{Riehle}, increasing the number of ions for example to
only 10 could shorten the integration time already by a full order
of magnitude. Together with improved laser stability and possibly
new detection methods, increasing the number of ions within a well
controlled ensemble of ions would substantially improve optical
frequency standards and open up ways for new applications
in navigation and geodesy~\cite{Bjerhammer}.

A pivotal question is, whether a larger sample of ions can be
controlled sufficiently well to reach fractional frequency
inaccuracies or long-term stabilities as low as $10^{-18}$ or
below. So far frequency standards based on microwave transitions
in a cloud of buffer gas cooled ions have been investigated for a
frequency standard with a long-term stability of
$10^{-15}$~\cite{Prestage}.  A ring of ions in an anharmonic
linear Paul trap has been proposed~\cite{Champenois} for an
optical frequency standard with improved short-term stability,
conserving the long-term stability at the level of $10^{-15}$.

For an optical ion clock at the accuracy level of $10^{-18}$, the choice of
atomic elements is limited as many ions have an electronic quadrupole moment $\theta$
in the excited state of the clock transition, which makes them
sensitive to electric field gradients.
In a linear ion trap with its large
static electric field gradient and the Coulomb fields of neighboring ions,
quadrupole shifts are on the order of 10 Hz. An exception are two-electron systems, where
transitions between $^1S_0$ and $^3P_0$ states with $\theta=0$ are
available. The most advanced ion clock today~\cite{Chou} uses such a system, $^{27}$Al$^+$, with $\theta=0$.
Here, the clock ion is stored in a linear Paul trap together with a second species
with easily accessible atomic transitions. The clock signal is read out via the second ion in
a quantum logic step~\cite{Schmidt,Rosenband2007}.
In general, the field of ion trapping has been pushed tremendously
by the developments in quantum computation with cold ions in the past few years. A high level control
of up to 14 ions has been demonstrated in entangled strings of
ions~\cite{BlattWineland,Blatt}, cooling to the three-dimensional ground state was
implemented~\cite{Monroe1995} and new scalable ion trap structures
are being developed for larger quantum
registers~\cite{Amini,Kielpinski}.

The most striking remaining problems for optical clocks are
heating of the ion due to its interaction with fluctuating patch
potentials~\cite{Turchette} and driven motion
(micromotion~\cite{Berkeland}) due to residual rf fields of the
trap, both giving rise to significant second-order Doppler shifts.
In the NIST Al/Mg ion trap a fractional frequency shift due to micromotion
of $3\times10^{-17}$ was observed, when the ion was shifted along
the axial trap direction by only 3 $\mu$m~\cite{Chou}.  This
excess micromotion (EMM) arises because of finite size effects,
misalignment and machining tolerances in the ion trap assembly,
see section 4, and poses severe problems, when scaling up an ion
optical frequency standard to many ions.

Inspired by scalable designs of ion traps for quantum
computation~\cite{Amini} we present a design study for an optical
frequency standard based on linear chains of ions with a potential
fractional frequency uncertainty of $10^{-18}$ or below.  In our
scalable, segmented four-layer Paul trap, chains of up to 10 ions
can be trapped in the Lamb-Dicke regime~\cite{Dicke} in each
trapping segment. It provides both, all degrees of freedom to
control micromotion as well as optical access in three dimensions.
In such a trap we plan to trap multiple chains of $^{115}$In$^+$
ions. The indium ion clock transition can be detected directly
using quantum jump spectroscopy. In order to ensure efficient
trapping and cooling of the ionic crystals, additional sympathetic
cooling with $^{172}$Yb$^+$ will be provided. The high precision
finite element method (FEM) calculations presented in this paper
are used to develop an ion trap made of AlN ceramics wafers in our
lab. In a prototype of this trap, based on a ceramic filled, glass
reinforced thermoset (Rogers4530B$^{\mathrm{TM}}$), we have
already successfully trapped chains of $^{172}$Yb$^+$ ions for
first tests on this design~\cite{tbp}.  Clearly, there are many
challenges ahead in building an optical clock  based on Coulomb
crystals with systematic frequency shifts controlled at the level
of $10^{-18}$ and issues of controlling the ion dynamics,
collisions with background gas, rf phase shifts, magnetic fields
and optical imaging will be addressed in our experiment with mixed
chains of ytterbium and indium ions in our scalable ion trap.

In summary, in this paper we propose a novel optical frequency
standard
based on linear crystals of $^{115}$In$^+$ ions sympathetically cooled by $^{172}$Yb$^+$.
We focus on the design of the ion trap for such an optical clock
with many ions. In particular, we address the problem of residual
micromotion and required machining tolerances.  As
micromotion causes extra heating in laser-cooled systems, this
study is of general interest for high precision experiments, for example
for investigating ion-atom interactions in degenerate quantum
gases~\cite{Zipkes}.

In section 2, we detail our approach to an
optical indium frequency standard with ionic crystals and show
that a long-term frequency stability of $10^{-18}$ or below with
systematic frequency shifts at the mHz level is possible. Section
3 gives general requirements and considerations regarding the ion
trap design. Section 4 describes the high precision FEM
calculations of the electric field of a segmented linear ion trap.
The influence of the geometry is discussed and critical tolerances are detailed.
\section{Systematic shifts in an optical indium multi-ion clock}
\label{sec:2}
\begin{figure}[hbtp]
\begin{center}
\includegraphics[width=0.45\textwidth]{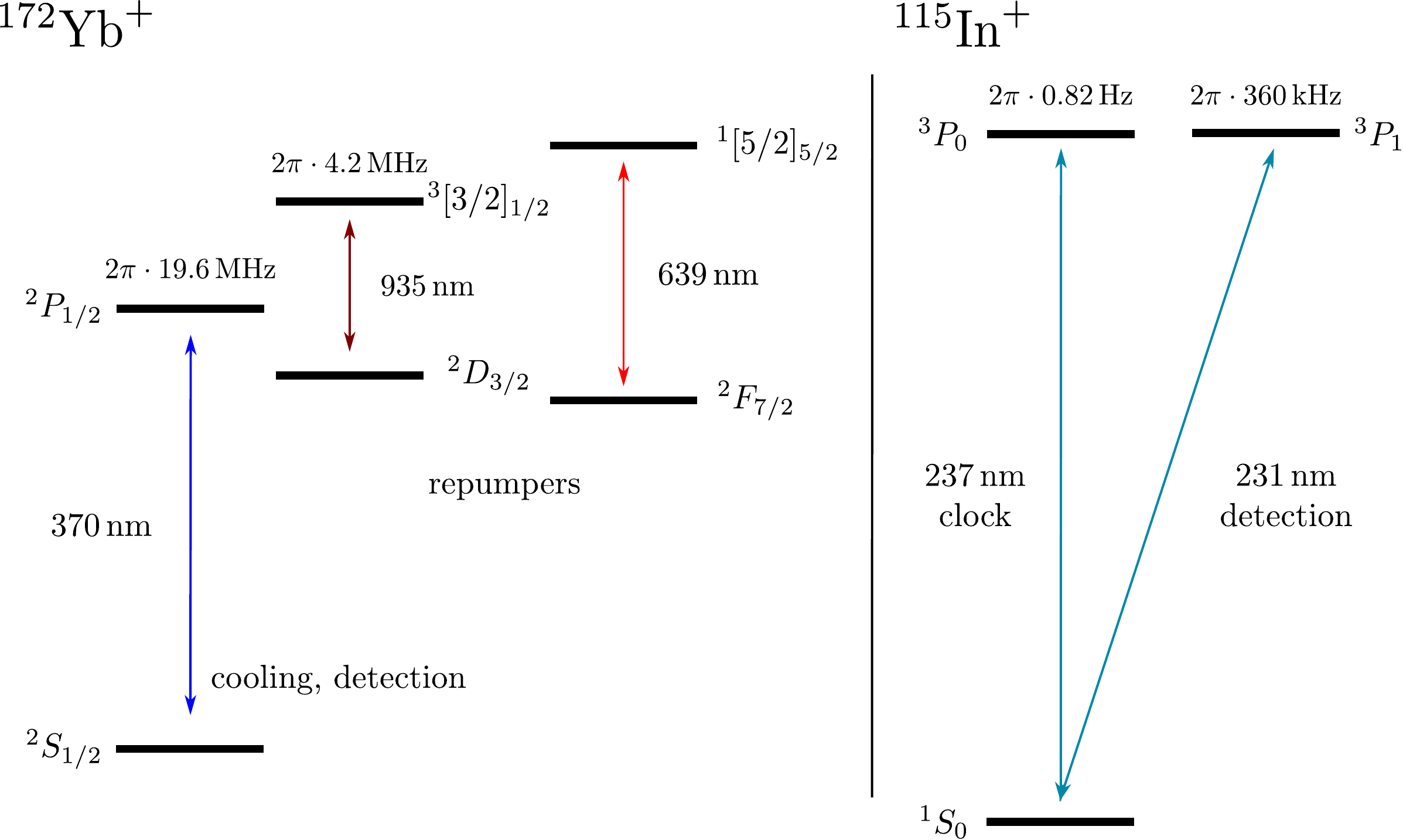}
\caption{Simplified level scheme of $^{115}$In$^+$ and
$^{172}$Yb$^+$, showing relevant cooling, repump and clock
transitions.}\label{term}
\end{center}
\end{figure}

The single $^{115}$In$^+$ ion is a well known, previously
investigated candidate for an ultra-stable and accurate optical
clock~\cite{Peik,Liu,Fortson}.  Its narrow clock transition
$^1$$S$$_0\to^3$$P$$_0$ at 236.5~nm, with a natural linewidth of
$\gamma=0.8$~Hz and electronic quadrupole moment $\theta = 0$, see
figure\ref{term}, makes it an interesting candidate for a scalable
optical clock with many ions.
The possibility to detect the quantum information of the clock
excitation directly via the $^3P_1$ state with a natural lifetime
of 0.44~$\mu$s ($\gamma=360$~kHz)~\cite{Peik} can facilitate the
atomic signal read-out of a larger chain of ions, without the need
of quantum logic technique~\cite{Schmidt}. Its transition
wavelength of 230.5~nm can be generated with standard diode laser
technology and second harmonic generation~\cite{Schwedes}. Due to
the lower cooling power obtained by laser cooling on this narrow
transition and the increased rf heating in Coulomb crystals, it is
still advisable to sympathetically cool the indium ions with a
second ion species with a stronger cooling transition. In our case
we choose $^{172}$Yb$^+$, because of its long lifetime in ion
traps, the all-diode-laser based easy accessible wavelengths and
the similar ion mass, that allows for efficient sympathetic
cooling. YbH$^+$ formed in collisions with background gas can be
dissociated with the Doppler cooling light~\cite{Sugiyama}. In
addition, the $^{172}$Yb$^+$ isotope has no hyperfine structure
and cooling is facilitated. Besides the enhanced cooling
efficiency and control of ions, the presence of a second species
allows for the possibility to sympathetically cool during the
clock interrogation in case of excessive heating, additional
characterization of the trap environment, such as magnetic fields,
and an alternative clock read-out via quantum logic for
comparative studies.

Besides the absence of electronic quadrupole moment, the $^1$$S$$_0\to^3$$P$$_0$
transition in $^{115}$In$^+$ has the advantage of a very low sensitivity to
environmental effects, which are summarized in table~\ref{tab:1}.
\begin{table}
\caption{Systematic relative frequency shifts in $^{115}$In$^+$.
For the
  uncertainty of the BBR shift the theoretical uncertainty is given.
  (*) a heating rate of 50 phonons per second at $\nu_{\mathrm{sec}}=1$~MHz is assumed, see
  text.}
\label{tab:1}       
\begin{tabular}{llll}
\hline\noalign{\smallskip}
Type of shift  &  Shift $\left|\Delta\nu\right|$  & Uncertainty  \\
  &  in $10^{-18}$ &  in $10^{-18}$  \\
\noalign{\smallskip}\hline\noalign{\smallskip}
Thermal motion (T=0.5mK) & 1 & 1   \\
Heating$^*$ & $<$ 0.1 & $<$ 0.1   \\
Excess micromotion & $\approx$ 1 & $<$ 1   \\
2nd order Zeeman & $<$ 0.1 & $<$ 0.1  \\
BBR at 300K & 13.6 & 1  \\
DC Stark & 0.08 & 0.08   \\
\noalign{\smallskip}\hline
\end{tabular}
\end{table}
In particular, indium profits from its heavy mass when considering
relativistic frequency shifts (second-order Doppler shift) due to
time dilation $\Delta\nu_{td}/\nu=-E_{\mathrm{kin}}/mc^2$, where
$E_{\mathrm{kin}}$ is the kinetic energy of the ion, $m$ its rest
mass, $c$ the speed of light and $\nu$ the frequency of the atomic
transition. At the Doppler cooling limit $T_D=0.5$ mK of
$^{172}$Yb$^+$ this frequency shift amounts to
$\Delta\nu_{\mathrm{td}}/\nu=-E_{\mathrm{kin}}/mc^2=-1 \times
10^{-18}$, where $E_{kin}=(5/2) \, k_\mathrm{B} T$ is the kinetic
energy due to thermal secular and micromotion in a linear ion
trap~\cite{Berkeland}. If relative uncertainties below $10^{-18}$
are targeted, further cooling such as sideband cooling of the
Yb$^+$ ion or direct cooling on the intercombination line of the
$^{115}$In$^+$ ion~\cite{Peik_1999} can be implemented, where
temperatures of $100$~$\mu$K can be reached. An additional kinetic
energy can be brought into the system by excess heating due to
thermally activated patch potentials and contaminations
on the electrodes~\cite{Deslauriers,Bergquist_Rosenband_Science}.
Assuming an enhanced electronic field noise with a power spectral
density of $S_E=10^{-12}$ (V/m)$^2$Hz$^{-1}$ in our ion trap with
electrode-to-ion distance of 0.7 mm, see figure 6
in~\cite{Epstein}, we obtain
a maximum relative frequency shift of $1 \times 10^{-19}$ due to
heating of the secular motion during the clock interrogation,
corresponding to 5 phonons at an eigenfrequency of the secular
motion $\nu_{\mathrm{sec}}$= 1 MHz. Here we assumed a maximum
clock interrogation time of 100 ms, limited by the natural
lifetime of the $^3P_0$ state of $^{115}$In$^+$. Commonly observed
heating rates in such ion traps can be a factor 10 to 100
lower~\cite{Epstein}, giving relative frequency shifts of
$10^{-20}$ and below.

Regarding ac Stark shifts due to blackbody radiation (BBR),
$^{115}$In$^+$ has one of the smallest sensitivities among optical
clock candidates. With a differential static polarizability of
$\Delta\alpha_0=5\times 10^{-8}$Hz/(V/m)$^2$ between the $^1S_0$
and $^3P_0$ state~\cite{Safronova}, it is comparable to
$^{27}$Al$^+$~\cite{Safronova,Rosenband_BBR}. Here, dynamic
corrections due to the frequency distribution are neglected, as they are typically of the
order of a few percent or below.  At room
temperature this corresponds to a blackbody shift of the clock
transition of $1.36(10) \times 10^{-17}$.  For example,
determining the temperature of the trap environment to $T = 30 \pm
10$~$^{\circ}$C will give a relative clock uncertainty due to the
blackbody shift of $2\times 10^{-18}$.

Two major systematic frequency shifts arising from imperfections
in the rf trapping potential are excess micromotion (EMM) and dc
Stark shifts. If the driving radiofrequency field
$E_{\mathrm{rf}}$ of the trap cannot be zeroed at the place of the
ion, it amounts to a second-order Doppler shift of
$\Delta\nu_{\mathrm{td}}/\nu= - e^2 E_{\mathrm{rf}}^2
/4c^2m^2\Omega_{\mathrm{rf}}^2$, where $e$ is the electronic
charge, $E_{\mathrm{rf}}$ the amplitude of the electric rf field
at the place of the ion and $\Omega_{\mathrm{rf}}/2\pi$ the trap
drive frequency~\cite{Berkeland}. This effect is the dominating
uncertainty in today's best ion clocks and aggravates for a larger
number of ions. Section 4 of this paper addresses the influence of
finite size effects of the trap and imperfections in the electrode
machining on the rf field in detail. The targeted upper limit in
our trap design for residual rf fields on the trap axis is a
maximum of 90 V/m, which corresponds to
$\Delta\nu_{\mathrm{td}}/\nu= -1 \times 10^{-18}$ for a trapped
$^{115}$In$^+$ ion at a trap frequency $\Omega_{\mathrm{rf}}/2\pi=
20$~MHz, as discussed in section 3.1. In addition, any residual electric
field seen by the ion, will give rise to a dc Stark shift $\nu_S$
in the clock transition.  Here, $^{115}$In$^+$ profits from its
low static differential polarizability
$\Delta\alpha_0$~\cite{Safronova}. At an rf field amplitude of 90
V/m this Stark shift corresponds to $\nu_\mathrm{S}=\Delta
\alpha_0 \left\langle E_{\mathrm{rf}}^2\right\rangle=8 \times
10^{-20} \times \nu$.

Lastly, we consider the influence of static and dynamic magnetic
fields in the ion trap.  The linear Zeeman shift from the
$m_\mathrm{F}=9/2$ to $m_\mathrm{F}=7/2$ states amounts to
6.36~kHz/mT~\cite{Becker} and is due to hyperfine mixing of
$^3P_0$ and $^1P_1$ states (the nuclear spin of $^{115}$In is
$I=9/2$). It can be measured and subtracted by alternatively
pumping the ion into the stretched states of the ground state with
opposite magnetic
moments~\cite{Madej,Bergquist_Rosenband_Science}. The second-order
Zeeman shift is given by $\Delta\nu=\beta \left\langle
B^2\right\rangle$, where for the alkaline-earth like system
$\beta=2 \mu_B^2/3h^2\Delta_{\mathrm{FS}}$ $=4.1$ Hz/mT$^2$, where
$\mu_B$ is the Bohr magneton and $h$ the Planck constant. Owing to
the large fine-structure splitting $\Delta_{\mathrm{FS}}=3.2\times
10^{13}$ Hz of the excited triplet states, indium has an
advantageously low second-order B-field dependency, 10000 times
lower than $^{171}$Yb$^+$ and 20 times lower than $^{27}$Al$^+$.
Due to unbalanced currents in the ion trap alternating magnetic
fields with $B_{\mathrm{rms}}^2= 2.2\times10^{-11}$T$^2$ have been
observed in ion traps~\cite{Chou}. For $^{27}$Al$^+$ this accounts
for an ac Zeeman shift of $\Delta\nu/\nu=1.4\times10^{-18}$. For
the $^{115}$In$^+$ ion with $\Delta\nu/\nu=7\times10^{-20}$ this
effect is negligible at the level of $10^{-18}$.  While static and
dynamic B-fields can be determined quite accurately and frequency
shifts can be taken account of, for an optical clock based on many
ions, requirements on the homogeneity of magnetic fields become
important.  With a linear Zeeman shift of 6.36 kHz/mT variations
in magnetic field amplitude should be less than 16 nT across the
ion chain to ensure that the broadening of the atomic line is less
than 0.1 Hz. Distortions of the line profile due to spatially
varying systematic shifts will have to be evaluated carefully to
avoid locking offsets. The fast interrogation cycle, that is
possible with a multi-ion frequency standard, will be advantageous
when both stretched states are probed, in order to avoid offsets
due to temporally varying magnetic fields.

The above considerations show that it should be possible to
evaluate the clock frequency of a larger sample of $^{115}$In$^+$
ions with a fractional frequency uncertainty of $10^{-18}$,
assuming that a sufficiently ideal trap can be machined, which
will be addressed in more detail in the following sections of this
paper. We have not addressed other potential systematic frequency
shifts like Doppler effects of
first-order~\cite{Bergquist_Rosenband_Science} or phase-chirps in
acousto-optic modulators, which are of general technical origin
and need to be evaluated in each individual experimental setup.
Frequency shifts due to collisions with the background gas have to
be studied, but are typically in the $10^{-19}$ range when
operating the ion clock at a background pressure in the low
$10^{-9}$~Pa range~\cite{Bergquist_Rosenband_Science}.

\section{Trap design}\label{trap_design}
\begin{figure}[hbtp]
\begin{center}
\includegraphics[width=0.4\textwidth]{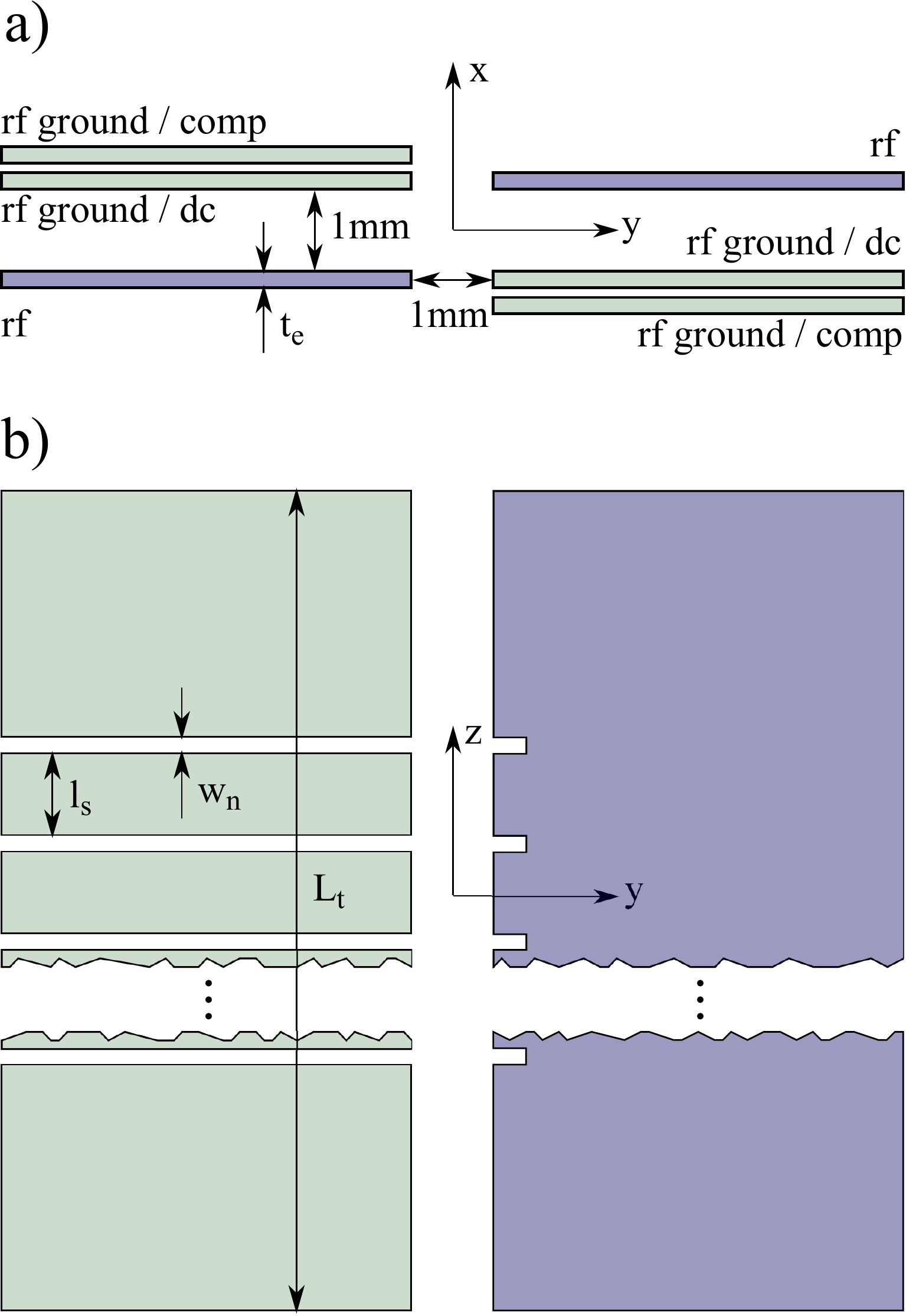}
\caption{Schematic drawing of our trap design based on four
segmented electrode wafers: (a) side view; (b) top view onto one
inner wafer with a rf trap electrode pair.
 The following lengths of the trap geometry are defined:
 electrode thickness $t_{\mathrm{e}}$; total length of the trap electrodes
 $L_{\mathrm{t}}$; segment length $l_{\mathrm{s}}$, and width of isolation
 slit in the rf ground / dc electrodes $w_{\mathrm{n}}$ which is identical
 to the width of the corresponding notches in the rf electrodes.}\label{roughtrapdesign}
\end{center}
\end{figure}

For an optical clock with low systematic frequency shifts and
competitive long term stability it is important to have a high
level control of the electrostatic and magnetic fields seen by
the ions over the whole trapping region. This includes both
necessary external control and 3D optical access to measure
micromotion and field shifts. We consider a scalable segmented
trap design, see fig.~\ref{roughtrapdesign}, in which ions can be
trapped either in a long linear string with an axial anharmonic dc
potential~\cite{Lin} or, in an harmonic axial confinement, in
separate ``buckets'' of up to 10 ions in every other trap segment,
that can then be controlled independently. Compared to linear ion
trap designs with separate dc electrodes on the trap axis for
the axial confinement \cite{Naegerl}, segmented rf
electrodes~\cite{Schulz,Madsen} give a more ideal behavior of the
rf field lines along the axial trap direction and
allow for a high precision in the alignment of trap electrodes, if
machined from one part. Currently a laser cutting and
metallization process for low rf loss ceramic wafers is being
developed in our lab. Four of these wafers can be precision
aligned and stacked on top of each other to give the trap
assembly.

Figure 1(b) shows the top view onto one pair of rf quadrupole
electrodes, that can be laser cut from one thin wafer. To the
segmented  rf ground electrodes different dc potentials can be applied
for the axial confinement as well
as  for static stray field compensation in the axial and
one radial direction. To obtain both degrees of freedom to
compensate for residual micromotion in the radial plane, our
design includes an additional electrode wafer with compensation
electrodes that overlie each rf ground/dc electrodes with a small
spacing in the range of 0.1~mm to 0.25~mm, see fig. 1(a). The dc
field generated by these extra compensation electrodes as well as
their effect on the rf field is described in
section~\ref{xtrcomps}. Our trap design is rather open and offers
optical access in the $(x, z)$ plane and in the $(y, z)$ plane for
laser beams necessary for laser cooling, exciting repumping
transitions, state manipulation and probing. In $x$
direction, with a typical electrode thickness $t_{\mathrm{e}}$
between 0.2~mm and 0.4~mm, a solid angle fraction larger than 4~\%
is available for optical fluorescence detection.

\subsection{Trap parameters}
In order to reduce heating of the secular motion of the ions we
have chosen a rather large distance $d=0.7$~mm from the electrodes
to the trap center, since observed excess heating rates scale as
$d^{-4}$~\cite{Deslauriers,Epstein}. The geometry of the linear
quadrupole electrode assembly has unity aspect ratio, with a distance of
1~mm between the tips of opposite electrodes, as shown in
figure~\ref{roughtrapdesign}(a). For this case, the quadrupole
term of the radial rf potential is only a factor 1.3
lower than in an ideal Paul trap with hyperbolically shaped rf
electrodes. In this sense, the trap geometry is relatively
efficient and sufficiently large trap frequencies of secular
motion in radial direction $\nu_r$ can be realized despite the
large distance to the trap electrodes. With an rf voltage
amplitude of 1.5~kV at an rf drive frequency
$\Omega_{\mathrm{rf}}/2\pi=20$~MHz we obtain trap frequencies for
the secular motion of $\nu_{r,\mathrm{Yb}}= 1.25$~MHz and
$\nu_{r,\mathrm{In}}=1.75$~MHz for $^{172}$Yb$^+$ and
$^{115}$In$^+$ ions, respectively.

An important criterion for first-order Doppler free clock
spectroscopy is the Lamb-Dicke condition with $\eta=k \Delta
x<1$~\cite{Dicke}. Here, $\Delta x$ is the rms spread of the
position of the ion and $k$ is the modulus of the wave vector of
the light exciting the transition. At an ion temperature
corresponding to the Doppler limit for Yb$^+$, $T_D=$0.5~mK, the
Lamb-Dicke parameter  in radial direction takes a value of $\eta=k
\Delta x= 0.45$ for the indium clock transition at 236.5~nm. For
the ground state of the harmonical oscillator, with
$\nu_{r,\mathrm{In}}=1.75$~MHz, $\eta = 0.13$.

The confinement in axial direction is generated by applying a dc
potential between neighboring segments in the rf ground
electrodes. For an harmonic confinement inside the Lamb-Dicke
regime, every other segment can be used as an ion trap with one segment for
dc confinement in between. We find that the curvature of the
axial trapping potentials has a shallow maximum as a function of
segment length $l_{\mathrm{s}}$ close to 1~mm dropping off by 10\%
at 0.8~mm and by 7\% at 1.2~mm. Regarding the number of traps
obtained per unit length of the quadrupole electrodes it seems
desirable to use a smaller segment length. However, for segment
lengths $l_{\mathrm{s}}$ smaller than 1~mm neighboring traps
become less well isolated from each other in the sense that
varying compensation voltages in one trap are sensed by
neighboring ion traps.
In addition, the depth of the static potential well on the trap
axis decreases monotonically with decreasing segment length
$l_{\mathrm{s}}$. As will be described in section~\ref{slits} the
choice of the segment length also influences the axial component
of the rf field, as far as the contribution from the notches and
isolation slits in the quadrupole electrodes is concerned. For our
design, we consider segment lengths in the range from 1~mm to
2~mm.

In an ideal linear Paul trap the rf field vanishes along the
symmetry axis of the ion trap and driven micromotion of the ion
can be minimized for a linear string of ions. To prevent
transitions of the linear chain of ions to zig-zag, helical and
more complex crystal structures, the ratio of radial to axial trap
frequencies of secular motion $\nu_r/\nu_z$ has to be kept above a
critical value, which depends on the number of ions $N$ in the
crystal. The various phase transitions in the ion Coulomb crystal
were investigated numerically and
analytically~\cite{Schiffer,Dubin,Totsuji}. Estimates for the
critical ratio $\nu_r/\nu_z$ for the transition from the linear
crystal as a function of $N$ from different authors are in
agreement with each other and are compiled in
reference~\cite{Wineland_Monroe}. Using the criterion $\nu_r/\nu_z
> 0.73 N^{0.86}$ obtained by Steane~\cite{Steane}, we find, for
example, $\nu_r/\nu_z> 3$ for five ions and $\nu_r/\nu_z> 5.3$ for
ten ions. Choosing 1.75~MHz for $\nu_r$ the linear confinement of
ten ions is possible for $\nu_z< 0.33$~MHz.

\subsection{Design criterion: micromotion in linear rf trap}

A principal limitation in today's optical ion clocks are frequency
shifts induced by excess micromotion~\cite{Berkeland}. The
observed EMM can originate from uncompensated static stray fields
that shift the ion off the nodal line of the rf field, residual
phase shifts of the rf potential on opposing electrodes and
imperfections in the ion trap geometry that lead to a non-vanishing
rf field along the trap axis. Our design allows to
compensate static stray fields for all degrees of freedom in the
radial plane to zero micromotion radially. RF phase shifts can be
suppressed by carefully designed electronic circuits, symmetric
leads and low resistivities. The electronic circuit scheme of our
scalable trap design is currently tested on a first prototype
trap~\cite{tbp}. In this paper we focus on the minimization of
residual rf fields $E_{\mathrm{rf,z}}$ along the trap axis, that
are due to the trap geometry and cannot be zeroed, thus severely
limiting the available trapping region, where micromotion is negligible.

The dominating frequency shift due to EMM is the second-order
Doppler shift, while dc Stark shifts induced by residual rf fields
are roughly an order of magnitude smaller in the case of ions with
low static polarizabilites, such as $^{115}$In$^+$ and
$^{27}$Al$^+$, see section 2. We therefore assume as a trap design
criterion a maximum second-order Doppler shift of $\left|\Delta
\nu_{td}/\nu\right|=10^{-18}$ due to residual axial rf fields.

As already pointed out in section 2, this relativistic frequency
shift contributes with $\Delta \nu_{td} / \nu = - E_{\mathrm{kin}}
/(mc^2)$ to the fractional uncertainty of the optical clock. Here,
the kinetic energy
$E_{\mathrm{kin}}=E_{\mathrm{kin}}^{(sec)}+E_{\mathrm{kin}}^{(mm)}$
of the ion of mass $m$ is given by the contributions
$E_{\mathrm{kin}}^{(sec)}$ from the secular motion and
$E_{\mathrm{kin}}^{(mm)}$ from the micromotion of the ion.
Assuming the secular motion to be thermalized at a temperature $T$
one obtains $E_{\mathrm{kin}}^{(sec)}=\frac{3}{2} k_{\mathrm{B}}
T$ for the three degrees of freedom of motion with
$k_{\mathrm{B}}$ denoting the Boltzmann constant. For the radial
degrees of freedom in a linear Paul trap, assuming the
confinement is given by a purely quadratic pseudopotential
\begin{equation}
\Psi = e^2 E_{\mathrm{rf}}^2/(4 m \Omega_{\mathrm{rf}}^2),
\end{equation}
 the thermal contribution of micromotion to the kinetic energy is equal to
the kinetic energy contribution from the secular motion
$E_{\mathrm{kin}}^{(mm)}=\frac{2}{2} k_{\mathrm{B}} T$, a result
obtained from both classical and quantum-mechanical treatment of
the ion motion~\cite{Wineland_Itano_1987}. This is the case when
the EMM is reduced to a negligibly small level in the radial
directions and when the contribution of the static potential to
the radial confinement can be neglected. The static potential has
a weakening effect on the radial confinement and thus results in
an increase of the micromotion contribution to the kinetic energy
which can be calculated using equation~10 in
reference~\cite{Berkeland}. In an ideal trap the axial confinement
is purely given by the static potential. The micromotion
contribution to the kinetic energy in axial direction can only
stem from excess micromotion due to a residual axial component of
the rf field $E_{\mathrm{rf,z}}$ along the trap axis. Neglecting
the contribution of the static potential to the radial confinement
we obtain
\begin{equation}
\frac{\Delta \nu_{\mathrm{td}}}{\nu}=-\frac{5}{2}
\frac{k_{\mathrm{B}} T}{m c^2}- \frac{e^2 E_{\mathrm{rf,z}}^2}{4
m^2 c^2 \Omega_{\mathrm{rf}}^2}, \label{tdformula}
\end{equation}
when the excess micromotion in radial directions is reduced to a
negligibly small level, which is possible using techniques
described by Berkeland~\emph{et~al.}~\cite{Berkeland} and taking
care in the design of the rf circuit to minimize rf phase shifts
between the quadrupole electrodes to a negligibly small level.

As the second term on the righthand side of
equation~\ref{tdformula} suggests, the axial component of the rf
field $E_{\mathrm{rf,z}}$ has to be considered separately.
Although a single ion or a linear chain of ions may also be
displaced in axial direction in the trap by applying a dc electric
field in search for a minimum in $|E_{\mathrm{rf,z}}|$, the
existence of a minimum and whether the minimal magnitude of
$E_{\mathrm{rf,z}}$ is small enough is determined almost entirely
by the design of the trap and the strict keeping of tolerances in
its construction. Hence, in the design of the trap we look for
geometries which minimize the axial component of the rf field and
analyze the sensitivity of different sizes in the geometry in
order to state the tolerances for the construction.

For indium ions using an rf drive frequency
$\Omega_{\mathrm{rf}}/2\pi = 20$~MHz and an rf amplitude of 1.5~kV
a residual axial rf field amplitude $|E_{\mathrm{rf,z}}|$ of
90~V/m amounts to a relative frequency shift of $\Delta
\nu_{td}/\nu=-1 \times 10^{-18}$ due to non-thermal EMM, second
term in eq.~\ref{tdformula}. In order to minimize the axial
micromotion for a linear chain of 5 to 10 ions to that level, in
each trapping segment a region of a length of 50~$\mu$m to
100~$\mu$m is required along the $z$ axis, where the magnitude of
the axial rf field is no larger than $90$~V/m.

\section{Trap field calculations}\label{trap_calculations}
An ideal infinitely long linear Paul trap has no axial rf field
component.  In segmented linear Paul traps of finite length used
in practice, a nonzero axial rf field component
$|E_{\mathrm{rf,z}}|$ on the trap axis appears for several
practical or technical reasons, which we investigated in order to
minimize their effect. These are: the finite length of the
quadrupole electrodes; insulation gaps in the dc electrodes; deviation from the ideally parallel
alignment of the quadrupole electrodes to the trap axis;
differences in the geometry of the electrodes and insulation gaps
that breach the symmetry.

For the trap field calculations in sections~\ref{finite_length}
through \ref{toltrans} an electrode geometry similar to
figure~\ref{roughtrapdesign} is used, with a varying number of
segments, as will be indicated in the text. Only the layers of
extra compensation electrodes are omitted, in order to preserve
symmetries and to separate the different effects. The effect of these
extra layers on the rf field is discussed in
section~\ref{xtrcomps}.

In order to obtain the ponderomotive potential given in equation 1,
the electric field $E_{\mathrm{rf}}$ of the linear Paul trap is calculated.
The Laplace equation is solved for boundary conditions given by the electrodes.
As this is a linear homogeneous differential equation, a
solution for a specific boundary condition can be written as a
linear combination of solutions of this geometry. It is worthwhile noting that a
better understanding of the effects can be gained by decomposing
the electrostatic potential problem linearly into one component
containing only the almost ideal part of the potential producing
the quadrupole trap potential and a second component, which
contains mainly the effect of the trap imperfection under
investigation. In our experiment we use a
configuration with two diagonally opposite electrodes put on rf
voltage $U_{\mathrm{rf}}$ keeping the other pair at rf ground. The
potential in this configuration is denoted by $(0,
U_{\mathrm{rf}})$. It can be thought of as sum of the
configurations $(-U_{\mathrm{rf}}/2, U_{\mathrm{rf}}/2)$ and
$(U_{\mathrm{rf}}/2, U_{\mathrm{rf}}/2)$. The configuration
$(-U_{\mathrm{rf}}/2, U_{\mathrm{rf}}/2)$ generates the quadrupole
potential required for trapping and by symmetry has zero voltage
everywhere on the trap axis. Therefore, for the ideal symmetric trap, this
configuration has no axial rf field component. The $(U_{\mathrm{rf}}/2,
U_{\mathrm{rf}}/2)$ configuration generates a flat nonzero
potential between the quadrupole electrodes with almost zero
electric field in the neighborhood of the trap axis.
It displays finite size effects and effects of the segmentation slits
onto the on-axis electric field clearly.

It can  also be advantageous to apply this decomposition method
when calculating the trap potential using the FEM.
The precision achieved at a given mesh point
density increases with decreasing electric field. Thus, effects
which appear mainly in the $(U_{\mathrm{rf}}/2,
U_{\mathrm{rf}}/2)$ configuration, which does not contain the strong radial
trapping field gradients, can be calculated with higher
accuracy. The dependence of the accuracy of the solution on the
electric field is accounted for by the definition of the residual
error estimation function~\cite{Goering_fem}, which plays an
important role in approximating iterative linear solver methods
employed in the solution of FEM problems as well as in adaptive
mesh refinement routines.

We estimate the accuracy of our calculations to resolve the axial
component of the rf field on the trap axis on a level well below
90~V/m at an rf amplitude $U_{\mathrm{rf}}$ of 1500~V. Regarding
the large radial field gradients of $2.3\times10^9$~V/m$^2$
obtained at this rf amplitude, this is demanding as the mesh
elements in the region of the trap axis typically have a size in
the range of 0.05~mm to 0.1~mm and hence, fields on their surfaces
are on average orders of magnitude larger.

To reach the required accuracy
we followed the mesh generation strategy with adaptive mesh refinement steps described in
Section~\ref{femcalcdetails} until a convergence at the low percent level is reached. Where applicable,
symmetry planes are used in the FEM model to reduce the size of the problem and hence required
computing power. For the trap model of section 4.2 and 4.3, the linear
decomposition into the configurations $(-U_{\mathrm{rf}}/2,
U_{\mathrm{rf}}/2)$ and $(U_{\mathrm{rf}}/2, U_{\mathrm{rf}}/2)$ could be exploited
and thus, higher accuracy reached at the same use of computing resources.
For these calculations we obtained accuracies better than 10~V/m for the axial
component of the rf field on the trap axis.

\subsection{Technical details of the finite element calculations}\label{femcalcdetails}
For the calculation of the rf trap potential, it is sufficient
to consider the rf field at a fixed phase, such that the field
calculation reduces to an electrostatic problem. Dynamic effects
due to the electronic lead design that can produce different phase
shifts for the rf field at different points in the trap have been
considered with a separate rf electronics software and will be
discussed for the prototype trap~\cite{tbp}.

The calculations of the rf field presented in this work consist in
finding the solution to the electrostatic Dirichlet problem, in
which the electrode surfaces of the trap are the boundaries kept
at given electric potentials. We use a commercially available
software (COMSOL Multiphysics~3.5) for finite element analysis. In
our FEM model we surround the trap electrode assembly with a wide
rectangular boundary box, which is set to ground potential. The
sides of the grounded box are typically at a distance of 1-2~cm
from the trap electrode surfaces. We verified for different box
sizes from 7 to 30 mm that this distance was large enough to have
no influence on the obtained results.

\begin{figure}[hbtp]
\begin{center}
\includegraphics[width=0.5\textwidth]{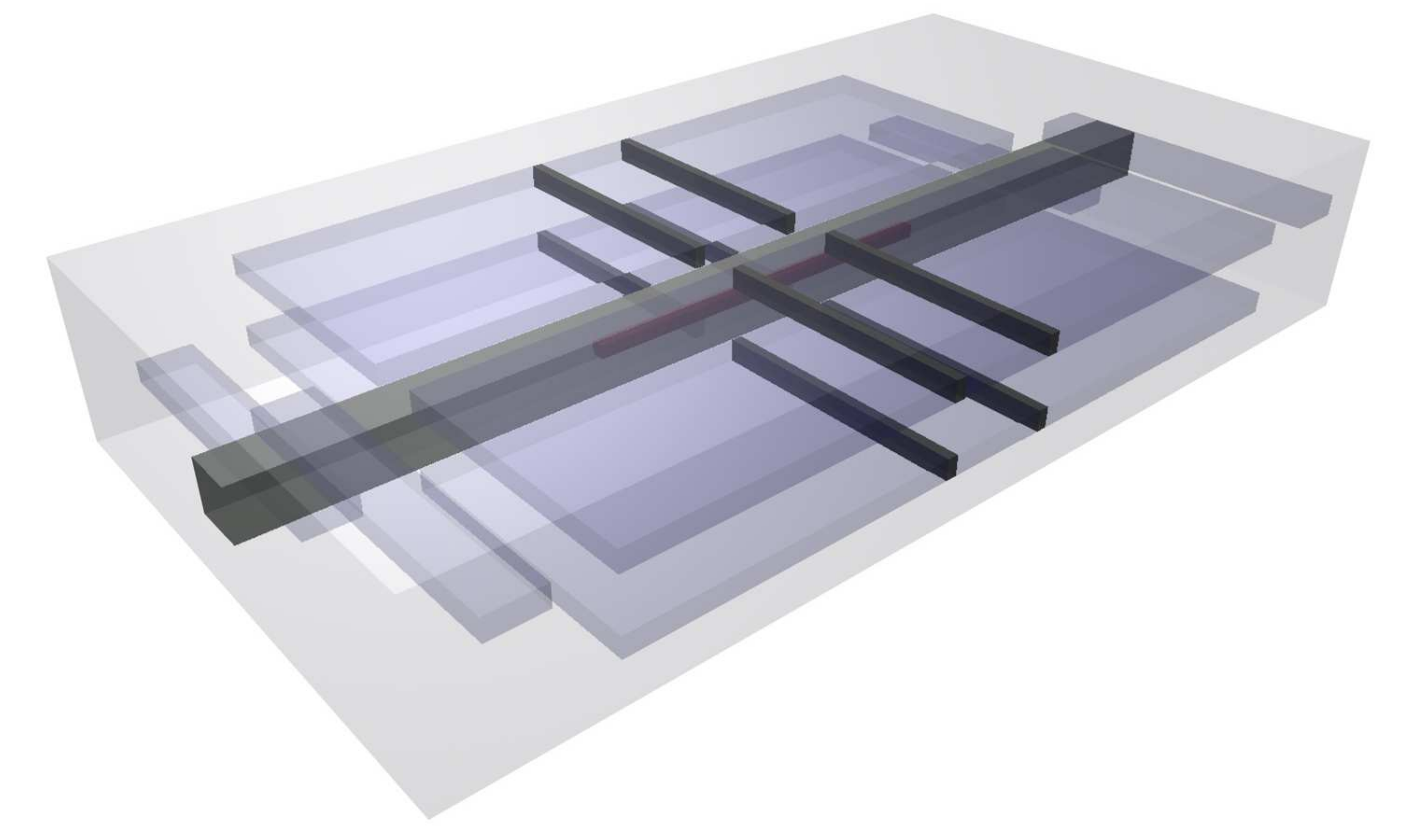}
\includegraphics[width=0.3\textwidth]{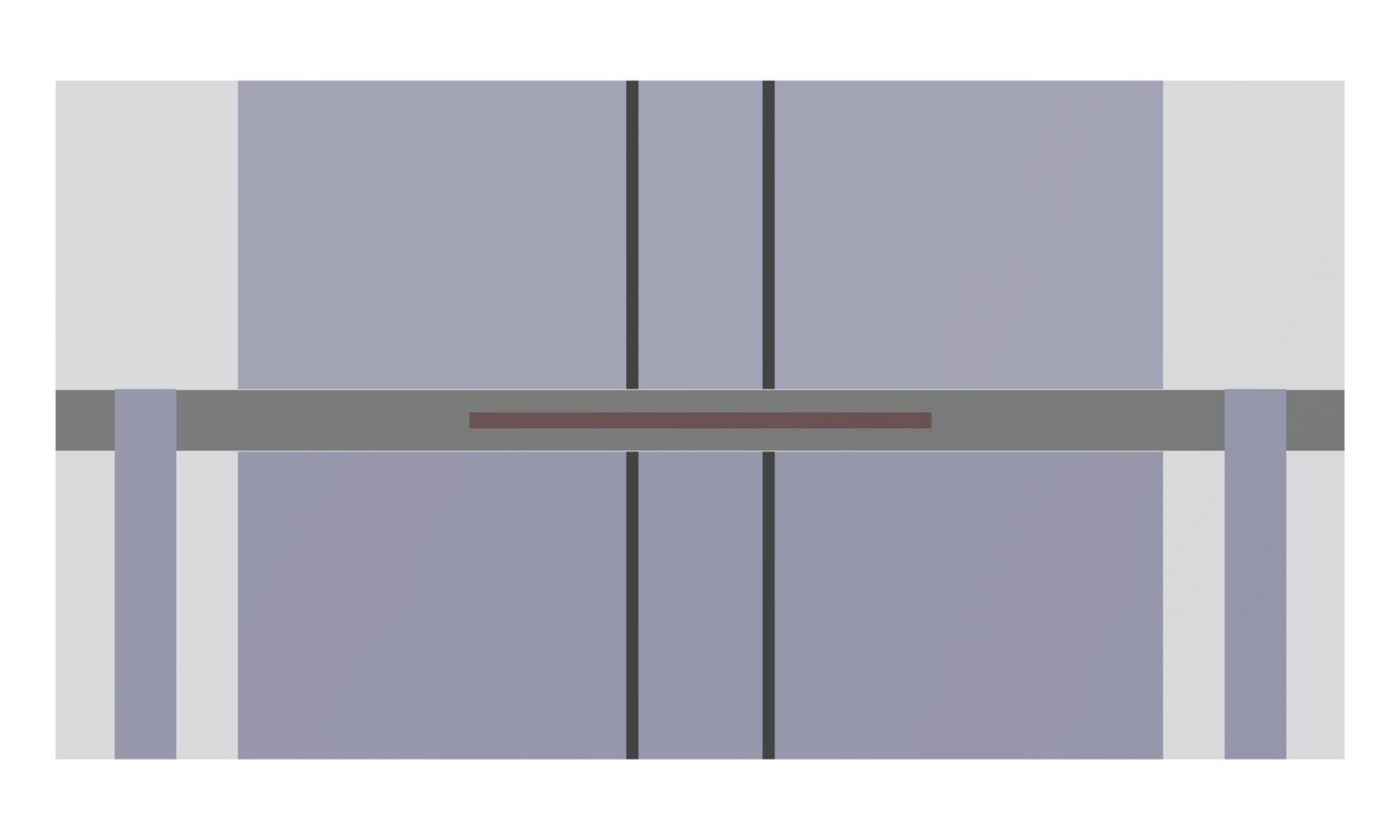}
\caption{(a) Schematic drawing of a geometry model with three
electrode segments used for FEM trap calculations. The trap
electrodes, including the grounded ends of the trap slit, are
displayed as light blue rectangular boxes. Dedicated domains for
mesh generation control are drawn in light to dark grey. The
outmost boundary box set to ground potential is not shown.  (b)
Top view showing electrodes and domains of the top layer with
grounded end electrodes.}\label{electrodescheme}
\end{center}
\end{figure}

Dielectric components required in the actual implementation of the
trap were not taken into account in our field calculations, since
they are either far away from the trap region or their surface is
almost entirely coated with conducting electrode material. The
electrode surfaces are represented in the geometry model of the
trap simulation by an assembly of rectangular boxes of 5~mm depth
and of variable width $l_s$ and thickness $t_e$
as then stated in the text. An example of a
geometric model used in a calculation
for a trap made of three segments is schematically drawn in
figure~\ref{electrodescheme}.

To generate the mesh we used the routines provided by the FEM
software. The mesh elements are of tetrahedral shape. Quadratic
Lagrange element functions were used in our calculations, with
the exception of the calculations presented in Fig.~\ref{notchpostol}
as here the use of quartic Lagrange element functions
led to a slight improvement in the quality of the solution.
As control parameters for the mesh generation we mainly used
the maximum mesh point separation and its growth rate in transitions
from regions where a higher mesh point density is required to regions
where it can be lower. In order to have an independent control
over the mesh generation in different regions of the geometry,
we define various domains where dedicated mesh generation parameters
can be applied (see Fig.~\ref{electrodescheme}).

An initial mesh is generated with the maximum mesh point separation
set equal to a fraction of the smallest distance in the surrounding
local geometry and the element growth rate to values between
1.1 and 2. We changed mesh control parameters in the special mesh
domains we had defined and compared the results of the calculation.
We found that choosing small element growth rates in the range
from 1.1 to 1.5, i.e. keeping mesh point density gradients small,
was important to obtain convergence of the result with increasing
mesh point density. Starting with a mesh, improved by the
convergence analysis with manually changing mesh parameters, we
additionally apply the adaptive mesh refinement routine of COMSOL.
During the solution of the problem this routine detects the
regions where improvements to the mesh are most effective in
increasing the accuracy of the calculation. It accordingly
improves the mesh quality and recalculates the solution. Typically
after two or three adaptive mesh refinement steps we found that,
regarding the axial component of the field on the trap axis, the
relative change between two consecutive adaptive mesh refinement
steps had decreased to a level on the order of $10^{-2}$.

The number of degrees of freedom or unknowns of the linear system
of equations for our problem was typically in the range between
$10^6$ and $8 \times 10^6$. Only approximating iterative solver
methods are practicable for problems of this size and the computer
memory at our disposal. In our case mostly an iterative conjugate
gradient method with algebraic multigrid preconditioning and an
iterative geometric multigrid method were used in combination with
the direct solvers PARDISO and UMFPACK. As computer hardware we
used two desktop computers, of which one has 4~gigabytes and the
other has 16~gigabytes of main memory.

\subsection{Effect of the finite length of quadrupole electrodes}\label{finite_length}
\begin{figure*}[hbtp]
\parbox{5cm}{\caption{(a) Axial field component of rf field along the trap axis for
  different length $L_t$, calculated for an
  electrode thickness $t_{\mathrm{e}}=0.2$~mm.  (b)~Width of range
  along trap axis, where the axial component $E_{\mathrm{rf, z}}$
  is of magnitude smaller than $90$~V$/$m as a function of
  the electrode length. Results for an electrode thickness
  $t_{\mathrm{e}}=0.2$~mm and for $t_{\mathrm{e}}=0.4$~mm are shown. Solid and dotted lines are to guide the eye.}
\label{graph_finitelength}} \parbox{13cm}{\begin{flushright}
\subfloat{\includegraphics[width=0.4387\textwidth]{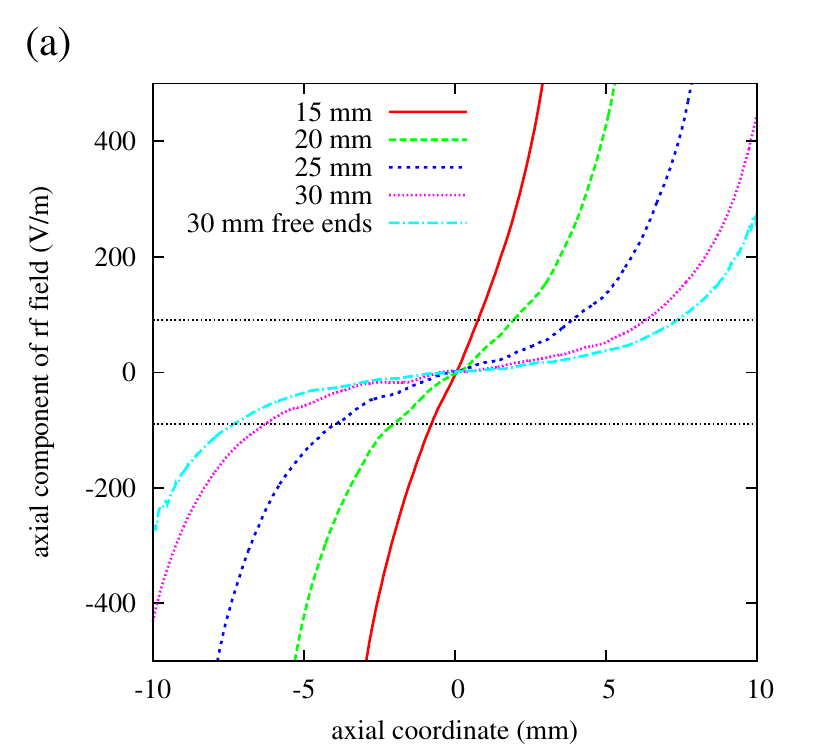}}
\subfloat{\includegraphics[width=0.2362\textwidth]{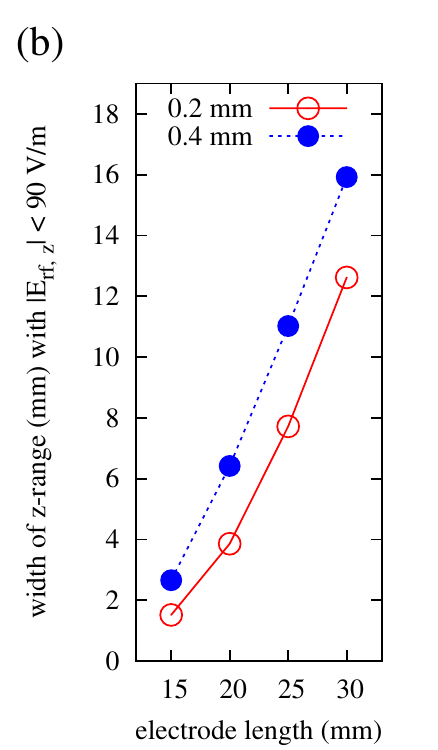}}
\end{flushright}}
\end{figure*}

For an electronically symmetric trap configuration \\
$(-U_{\mathrm{rf}}/2,U_{\mathrm{rf}}/2)$ the electrostatic
potential along the trap axis is 0. Therefore, the electric field
lines show no axial field component along the entire trap axis.
For an highly segmented rf trap, where trapping electrodes are
used simultaneously for dc compensation, it is technically very
challenging to deliver multiple independent rf voltages to the
various electrode segments with identical amplitude and phase.
Therefore, for the scalable trap design we chose the
$(0,U_{\mathrm{rf}})$ configuration, where one rf trap electrode
is kept on rf ground, facilitating independent adding of dc
voltages. In this configuration the electrostatic potential on the
trap axis is roughly $U_{\mathrm{rf}}/2$ and electric field
gradients along the axis are observed when terminating the rf trap
electrodes to a finite size. In order to keep this axial component
of the rf field amplitude below a given value, we adjust the
minimum length of the trapping electrodes. For our linear Paul
trap with a 1~mm~$\times$1~mm electrode spacing,
figure~\ref{graph_finitelength}(a) shows the axial component of
the rf field along the trap axis plotted over the axial coordinate
$z$ for various lengths of the quadrupole electrodes. Here, the
electrode thickness $t_e=0.2$~mm and, for simplicity, an electrode
geometry without notches was used.
Our trap design also includes grounded electrodes at the ends of
the quadrupole electrodes perpendicular to the trap axis behind a
1~mm insulating slit, as shown in figure~\ref{electrodescheme}.
The purpose of the grounded end electrodes is to provide a defined
electrostatic potential between the ends of the quadrupole
electrodes of the trap, where otherwise the insulator material of
the wafer would be exposed. For the calculations shown in
figure~\ref{graph_finitelength}(a) the grounded end electrodes are
included in the trap geometry. For comparison, for an electrode
length of 30~mm both cases are shown, with grounded and without
grounded electrodes (free ends curve).

Requiring an axial rf field component $E_{\mathrm{rf, z}}$ of
magnitude smaller than $90$~V$/$m, we
can find the width of the range on the trap axis usable for
trapping of clock ions.  In figure~\ref{graph_finitelength}(b) the
width of this usable range is plotted as a function of the length
of the quadrupole electrodes. When we segment the quadrupole
electrodes using a segment length of 1~mm, for example, with a
total length of the quadrupole electrodes of $L_t=30$~mm and an
electrode thickness of $t_e=0.2$~mm, our design can accommodate a
linear array of six usable ion traps for the clock application.
The number of usable traps decreases rapidly when the total
electrode length is shortened. With $L_t=2$~cm and $t_e=0.2$~mm
the number of usable ion traps is already reduced to two. For an
electrode thickness of 0.4~mm the corresponding usable range for
clock application is given in figure~\ref{graph_finitelength}(b)
as well. In comparison to $t_e=0.2$~mm, we obtain a moderately
increased number of suitable traps: eight with an electrode length
of 30~mm and three with a 20~mm electrode length.

\begin{figure}[hbtp]
\begin{center}
\includegraphics[width=0.45\textwidth]{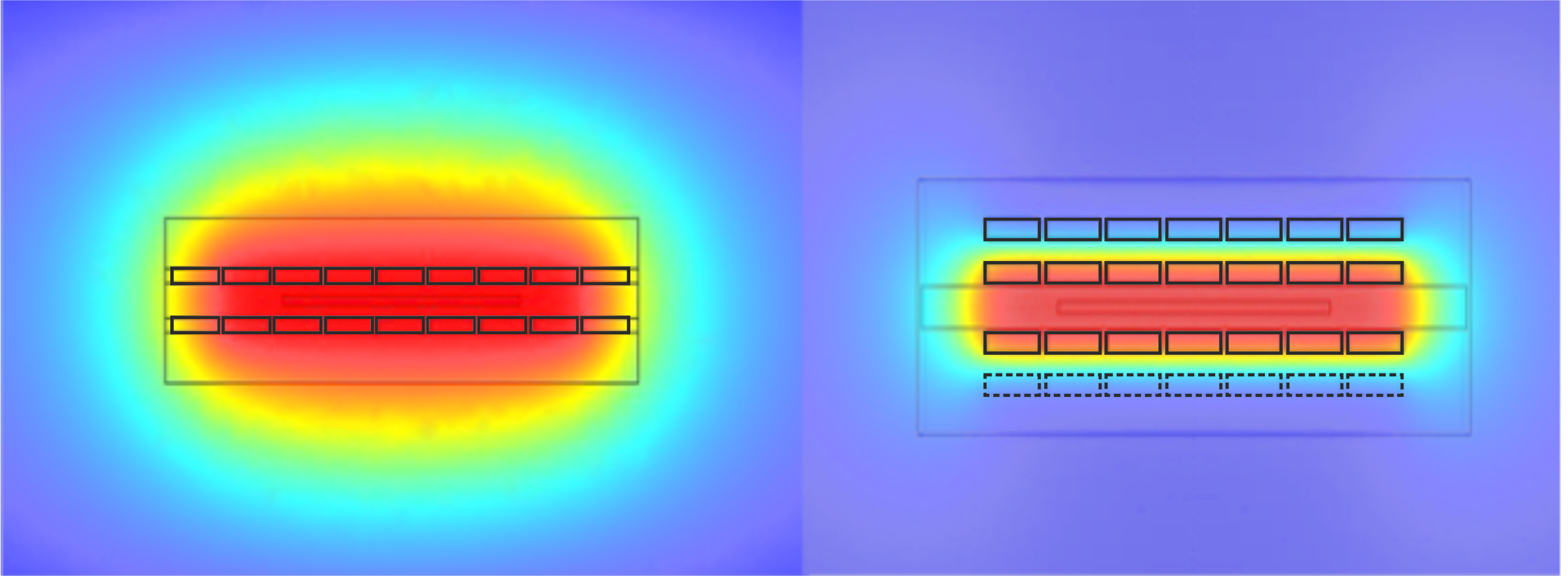}
\caption{False-color representation of rf equipotential surfaces of a
linear quadrupole trap.
(left) RF potential of a two-layer trap; (right) rf potential with additional rf ground electrode
layers ontop of rf electrodes.}\label{threelayerrfpotflat}
\end{center}
\end{figure}

With a view to further miniaturization of trap assemblies and
further increase of the number of trap sites, it would be
desirable to find a way to avoid long trapping electrodes. By
introducing an additional rf ground electrode above and below the
rf electrodes the effect of finite length of the trap electrodes
can be decreased substantially, such that more trap segments can
be used with the same or an even smaller length of trap
electrodes. Enclosing the linear quadrupole electrodes by extra
ground electrode layers from above and below forces
equipotential surfaces of the rf potential to
flatten out over a wide range of the trapping
area, as shown in figure~\ref{threelayerrfpotflat}~(right). With
electric field lines being orthogonal on equipotential surfaces,
this results in a smaller gradient of $E_{\mathrm{rf,z}}$ over
much of the $z$ axis, with a steeper increase only close to the ends of the trap
array.
However, there are several drawbacks of the additional rf ground
electrode layers, which are more or less important depending on
the choice of electrode thickness and electrode layer spacing: the
capacitance of the trap is increased; the numerical aperture for
optical detection can become limited by the electrode assembly;
increased risk of electrical breakdown; and finally, the axial rf
field component on the trap axis resulting from notches in the
electrodes is considerably intensified. As operating a 30 mm long
ion trap in our setup brings no further limitation for the moment,
we did not include this option in our current trap design.
Additional electrode layers are chosen only for extra compensation
on top of the rf ground electrodes as shown in
figure~\ref{roughtrapdesign}, that can then be put substantially
closer.

It is worthwhile mentioning that $E_{\mathrm{rf, z}}$ plotted in
figure~\ref{graph_finitelength}(a) can be well approximated in the
central region on the $z$ axis using the expression
\begin{equation}
f(z)=\frac{C}{\sqrt{r^2+(L/2+z)^2}}-\frac{C}{\sqrt{r^2+(L/2-z)^2}},
\end{equation}
which gives the axial component of the field of a line charge of
length $L$ in the distance $r$ from the $z$ axis and with unit
line charge density. For $z$ values closer to the ends of the
quadrupole electrodes the calculated field component
$E_{\mathrm{rf, z}}$ deviates strongly from this expression as
here the approximation using a constant line charge density can
obviously not be made. Using a least square fit of this expression
to the FEM calculation results for the axial electric field we
obtain values for $L$, $r$ and a proportionality factor $C$ for
different lengths of the quadrupole electrodes. In the
calculations presented in the following paragraphs it was helpful
to use a smaller length of the trap electrodes in order to keep
the size of the problem smaller. To better isolate the effects
under investigation we subtracted the finite length effect in form
of the mentioned fit function from the FEM results.
\subsection{Effect of insulation slits in dc electrodes}\label{slits}

In our design the rf ground electrodes also serve to apply dc
voltages for axial confinement of the ions and to generate
electric compensation fields in axial and in one radial direction.
The electronic insulation of the electrodes requires segmentation
slits. Here a notch is machined into the underlying insulator
material and a conductive layer is applied around the teeth
structure to cover insulation material in direct view of the ions,
still leaving each tooth electrically insulated from its
neighbors.

\begin{figure*}[htbp]
\parbox{5cm}{\caption{(a) Axial field component of rf field on the trap axis for
  different widths $w_n$ of the notches in the quadrupole electrodes. The
  electrodes are notched at $z=0$. The horizontal lines
  indicate the targeted maximum amplitude of 90 V/m.
  (b)~Maximum magnitude of the axial component of the rf field
  $|E_{\mathrm{rf, z}}|$ as a function of $w_n$. Lines between data points are to guide the eye.}
  \label{notcheffect}} \parbox{13cm}{ \begin{flushright}
\subfloat{\includegraphics[width=0.4387\textwidth]{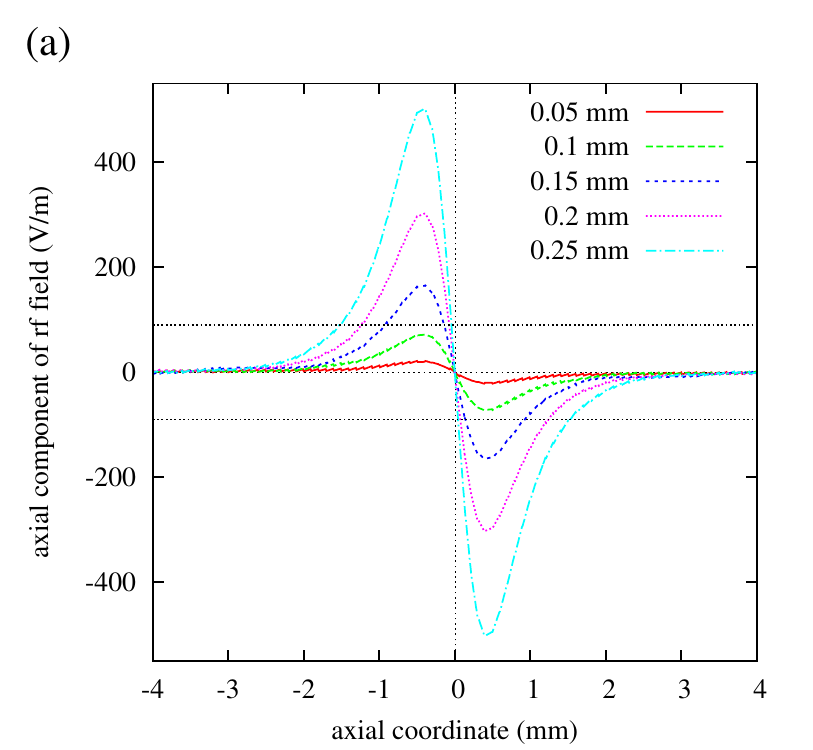}}
\subfloat{\includegraphics[width=0.2362\textwidth]{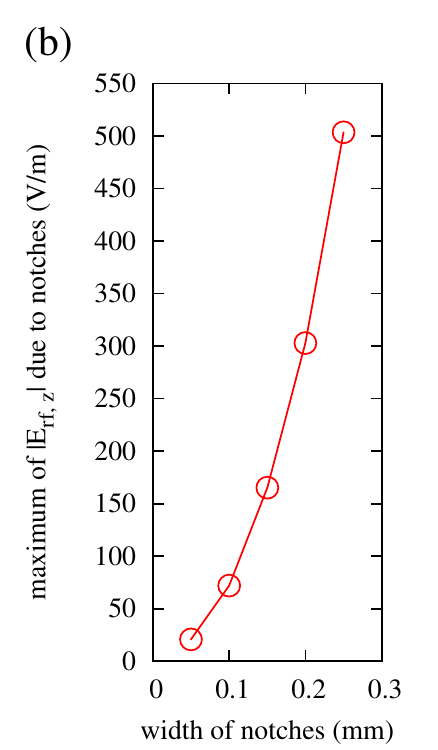}}
\end{flushright}}
\end{figure*}

Electrode surfaces of the trap assembly are equipotential surfaces
of the electrostatic problem. The introduction of insulation slits
into the electrode geometry causes a receding of equipotential
surfaces in their vicinity.  This receding is of course more
prominent closer to the electrode. Further away, it is
increasingly smoothed out and the curvature it causes in the
equipotential surfaces is reduced. Still, in the neighborhood of
the trap axis a net effect of insulation slits in the electrodes
is noticeable. Implementing corresponding symmetric notches in the
rf electrodes can partially compensate for this effect. This can be
understood when decomposing the corresponding electrostatic
problem linearly into the two configurations $(-U_{\mathrm{rf}}/2,
U_{\mathrm{rf}}/2)$ and $(U_{\mathrm{rf}}/2, U_{\mathrm{rf}}/2)$.
For the contribution of the $(-U_{\mathrm{rf}}/2,
U_{\mathrm{rf}}/2)$ configuration it is clear that keeping an
exact geometric symmetry between the rf and rf ground electrodes
leads to a zero potential all along the trap axis. On the trap
axis, the effect of notches in the rf ground electrodes is then
exactly cancelled out by the effect of the corresponding notches
in the rf electrodes. However, even if the symmetry of the
geometry is intact, there will be a residual magnitude in
the axial component of the rf field which stems from the
contribution of the $(U_{\mathrm{rf}}/2, U_{\mathrm{rf}}/2)$
configuration and can not be compensated by symmetry. The
potential on the trap axis in this configuration is a certain
amount smaller than $U_{\mathrm{rf}}/2$, depending on how open the
electrode structure is. For trap electrodes, for example, being
further apart from each other, the electrostatic potential on the
trap axis is smaller. This seems to be exactly the effect of the
notches in the electrodes, the equipotential surfaces are receded
closer to the notches, leading to a minimum in the rf
potential on the trap axis and resulting in a dispersion-shaped
behavior of the axial component of the rf field.
Figure~\ref{notcheffect} shows the calculation of this effect for
different widths $w_n$ of the notches in the electrodes.
In this calculation a thickness $t_e$ of 0.2~mm was assumed for
the electrode blades. To highlight the effect of a single
segmentation, a trap geometry with only one notch per electrode
blade was simulated. All four notches of the quadrupole electrodes
are located at $z=0$.

\begin{figure}[htbp]
\begin{center}
\includegraphics[width=0.45\textwidth]{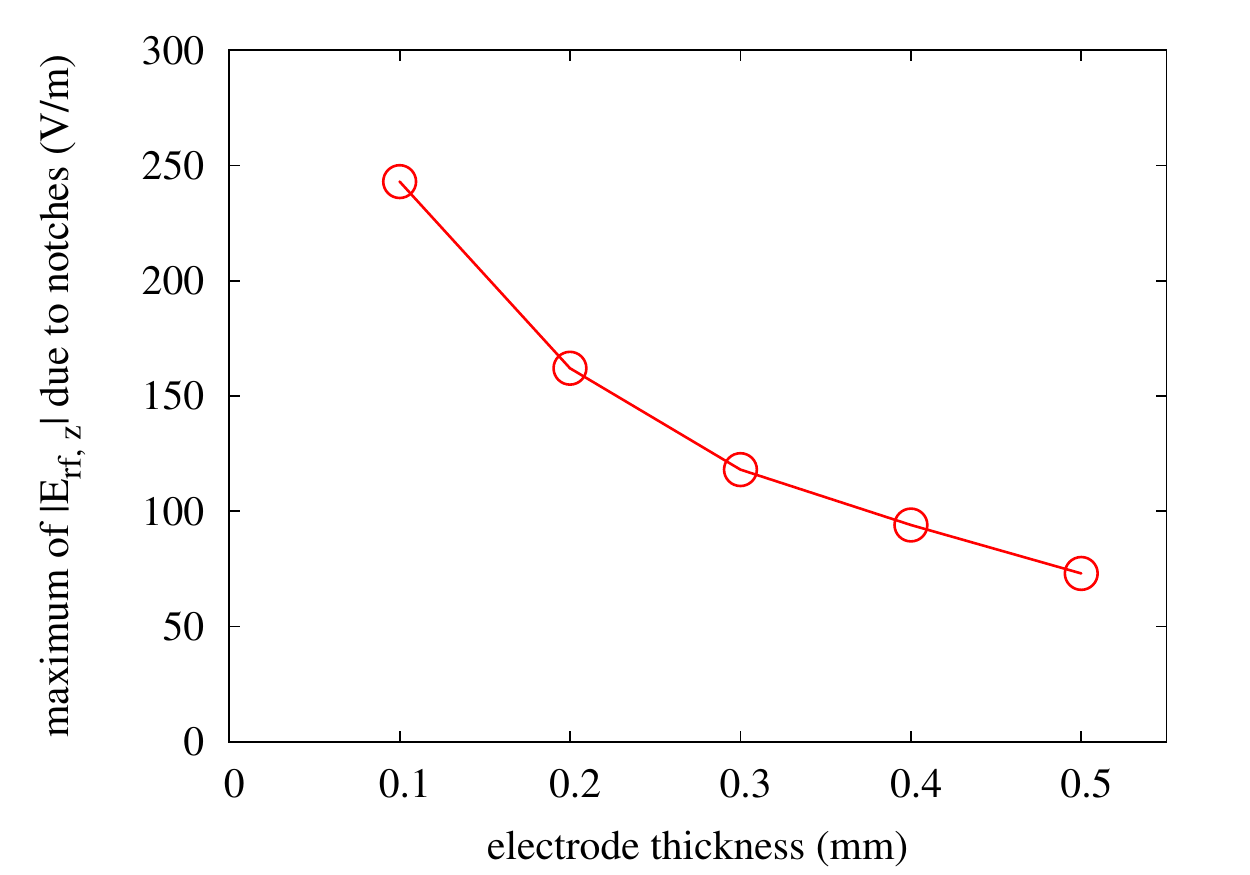}
\caption{Maximum magnitude of the axial component of the rf
  field $|E_{\mathrm{rf, z}}|$ as a function of the electrode thickness
  $t_e$ for notches of size $w_n= 0.15$~mm. Lines between data points are to guide the eye.}
\label{ezmax_elthickness}
\end{center}
\end{figure}

We also calculated the effect for different electrode thickness.
The maximum magnitude of the axial component of the rf field
$|E_{\mathrm{rf, z}}|$ obtained for different electrode thickness
$t_e$  is given in
figure~\ref{ezmax_elthickness} for a fixed notch size $w_n=0.15$~mm.
Using thicker electrodes is obviously advantageous regarding the
contribution of the notches to the axial component of the rf
field. This can be an interesting option when the electrodes are
made using a technique different from the one we chose. Both laser
cutting of thin wafers and the subsequent metallization are more
difficult for thicker structures and higher aspect ratios.

\begin{figure}
\begin{center}
\includegraphics[width=0.4\textwidth]{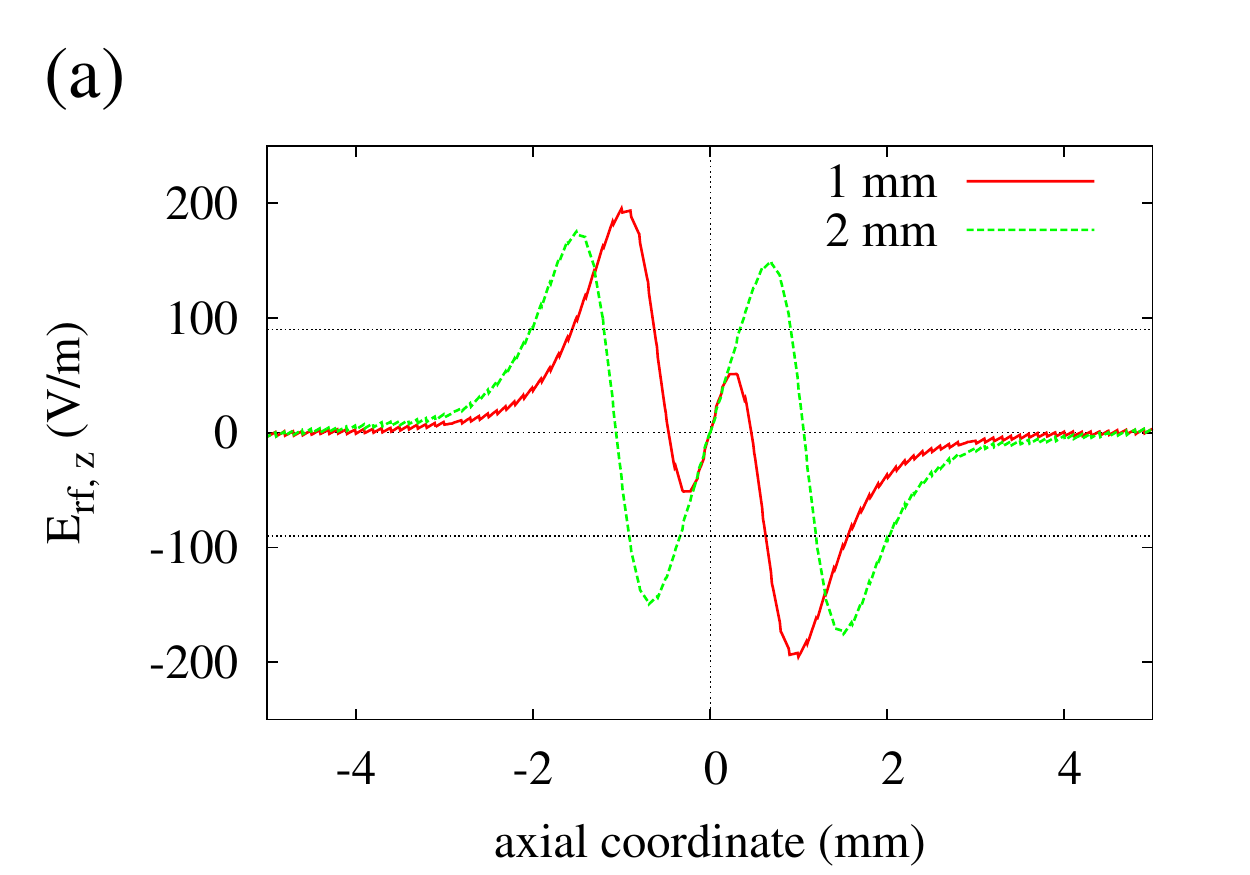}
\includegraphics[width=0.4\textwidth]{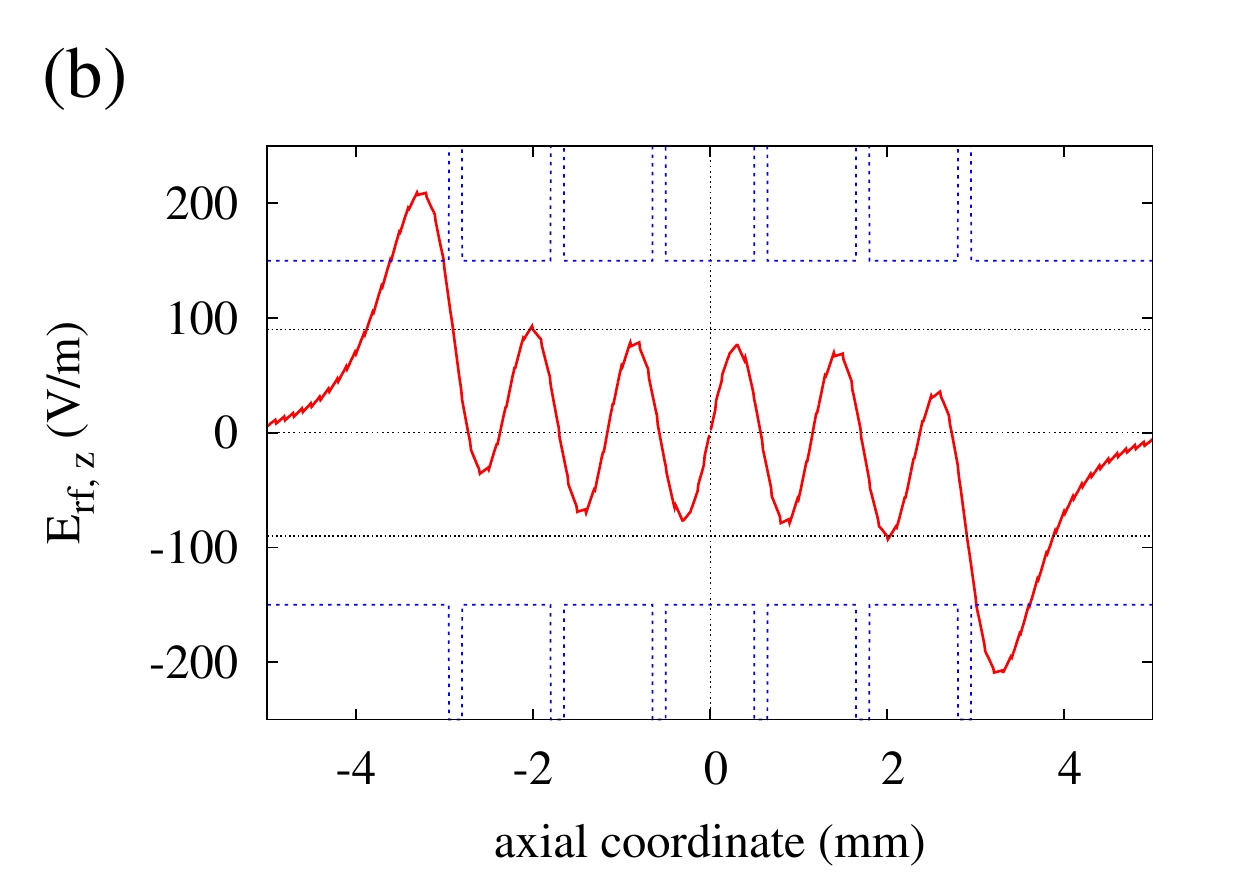}
\caption{(a) Axial component of the rf field $E_{\mathrm{rf, z}}$
on axis as a function of the axial coordinate $z$ in a single trap
for a segment length $l_{\mathrm{s}}=1$~mm and for a segment
length $l_{\mathrm{s}}=2$~mm with the width of the
  notches being 0.15~mm. (b) Axial component of the rf field
  $E_{\mathrm{rf, z}}$ on axis as a function of the axial coordinate
  in an array of traps with 5 inner segments of length $l_s=1$~mm
  separated by 0.15~mm wide notches. The axial location of the
  segments is indicated by dotted lines.}
\label{doublslseglen}
\end{center}
\end{figure}

For a segment length of a few mm and smaller, the effects of the
notches from both ends of a segment cancel partially. The behavior
of the resulting axial rf field in the middle of the segment
varies with the segment length, since it is composed of
contributions from both ends of the segment, which have opposite
sign. Figure~\ref{doublslseglen}(a) shows the simulation of a trap
with three segments, thus two discontinuities in the electrode
blades. The inner segment length is $l_s=1$~mm and $l_s=2$~mm,
respectively, with the width of the notches kept at 0.15~mm and
electrode thickness at 0.2~mm. Here, the center of the inner
segment is at $z=0$.
The curve for the 1~mm long segment shows, that the effects of the
notches from both sides of the segment are partially cancelling
each other. This can also be seen in the calculation for a trap
with five inner segments of 1~mm length shown in
figure~\ref{doublslseglen}(b). We can conclude, that if the
symmetry of the electrode construction is given, the remaining
axial rf field component produced by the notches in the electrodes
is small enough to allow trapping of ion chains of five to ten
ions for clock applications. Even if a larger width of the notches
is chosen, close to the center of each trap segment the axial rf
field component has a zero crossing resulting in a range with
$|E_{\mathrm{rf, z}}|<90$~V/m large enough to accommodate a chain
of ions.

\subsection{Angular misalignment of quadrupole electrodes}\label{tolangle}

Due to the precision achieved in the process of laser machining of
1 or 2 $\mu$m, we can assume that electrodes located on one wafer
can be aligned nearly perfectly with respect to each other. Thus,
we only need to consider angular misalignment of the two
quadrupole electrode pairs occurring during the trap assembly.

\begin{figure*}[htbp]
\parbox{5cm}{\caption{(a)~Axial component of the rf field
  $E_{\mathrm{rf, z'}}$ on the trap axis $(z')$ as a function of the axial
  coordinate for different misalignment angles $\alpha$. Here the trap
  axis $(z')$ is rotated by the angles $\alpha/2$ and $-\alpha/2$ to the
  axes defined by the upper and lower electrode layer,
  respectively. (b)~Magnitude of the axial component of the rf field
  $E_{\mathrm{rf,z'}}$ on the trap axis as a function of the
  misalignment angle. Data points are fitted with a linear fit to extrapolate
  $E_{\mathrm{rf,z'}}$ to small angles.}\label{normalangletol}} \parbox{13cm}{\begin{flushright}
 \subfloat{\includegraphics[width=0.4387\textwidth]{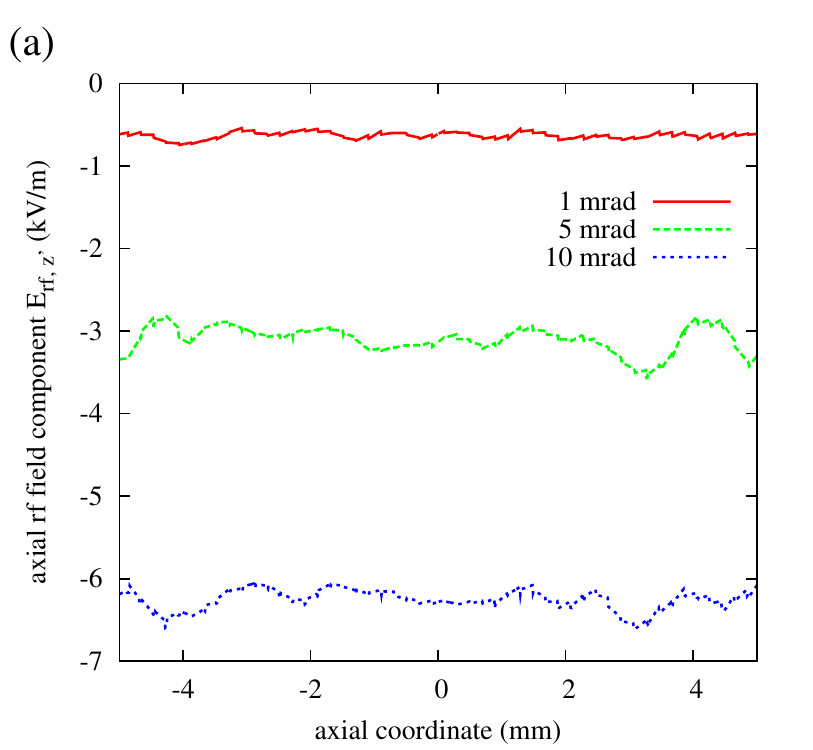}}
 \subfloat{\includegraphics[width=0.2362\textwidth]{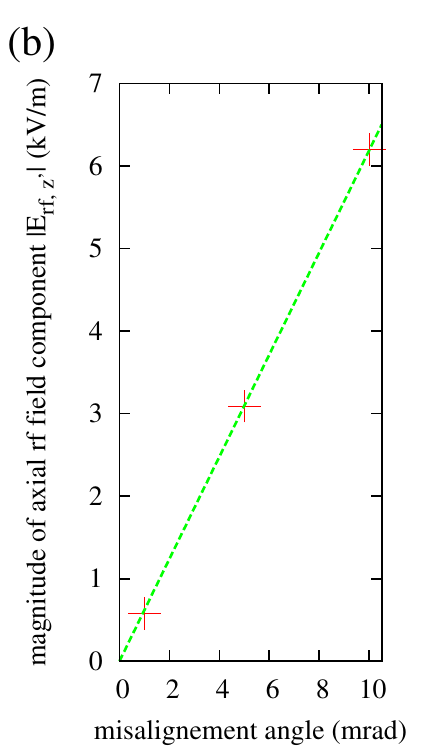}}
\end{flushright}}
\end{figure*}

First, we consider a possible rotation angle $\alpha$ between the
two wafers of the quadrupole trap electrodes around the axis
normal to the wafer plane, corresponding to the $x$ axis in
figure~\ref{roughtrapdesign}. Figure~\ref{normalangletol}(a) shows
the axial component of the rf field calculated for an electrode
thickness of 0.2~mm for different misalignment angles $\alpha$ using a $(0,
U_{\mathrm{rf}})$ configuration. Here, the trap
  axis $z'$ is rotated by the angle $\alpha/2$.
Our calculations show that this angular misalignment produces an
rf field component along the trap axis, which is proportional to
the angle of rotation of the two electrode layers with respect to
each other. This can be seen from figure~\ref{normalangletol}(b)
which shows a plot of $E_{\mathrm{rf,z'}}$ as a function of the
rotation angle. When repeating the calculation for an electrode
blade thickness of 0.4~mm the results for the axial rf field
component show no significant difference compared to the results
obtained with 0.2~mm electrode thickness. We deduce from
figure~\ref{normalangletol}(b) an axial rf field of $6.2 \times
10^2$~V/m per mrad.  In order to keep this axial rf field
component well below 90~V/m we have to ensure that the angular
misalignment of the two wafers for rotation around the $x$-axis is
less than 0.14~mrad.

Second, we consider an angular misalignment of the two electrode
wafers with respect to each other around the axis of rotation
corresponding to the $y$ axis (see fig.~\ref{roughtrapdesign}). An
angular misalignment around this axis can occur during the
assembly of the electrode wafers, when, for example, the spacer
height is not identical, or glue is creeping between spacer and
electrode blade. Such a misalignment leads in first order to a
linearly changing distance between the electrode wafers along the
trap axis. This obviously leads to a corresponding change in the
radial rf field gradients and consequently of the radial trap
frequencies for the secular motion and a small variation of the
on-axis potential is to be expected as well. In our calculations
we used misalignment angles in a range from 1~mrad to 10~mrad.
However, within the estimated accuracy of the calculation, which
is well below 90~V/m on the trap axis, we could not resolve any
deviation from zero in the axial rf field component due to angular
misalignment of this type.

\subsection{Tolerances on width and position of notches in electrodes}\label{tolslitwidthpos}

In the process of machining notches into the electrode blades
inaccuracies can occur in both the positioning of the notches as
well as in their width. The subsequent process of coating the
insulator with the electrode material can also contribute to
further inaccuracies.

First, we consider the effect of a small change in the width of
one of the four notches separating two segments of the linear Paul
trap. We calculated the resulting rf field when one notch is made
wider in one rf carrying electrode. Our calculations show what can
be expected from such a change in the geometry: an additional
receding of the equipotential surfaces away from the widened notch
occurs, centered at the axial coordinate of the notch.  As a
result the axial rf field component displays a dispersion-shaped
behavior along the $z$-axis at the edge of the trap segment in
excess of the magnitude found for a symmetric electrode geometry
 as shown in figure~\ref{notcheffect}(a). The radial components of the
rf field as well display on the $z$-axis an increase from zero
which peaks at the $z$ location of the widened notch. The feature
in the radial field components is of similar width and amplitude
as the one in the axial component. Due to the strong gradient in
the radial rf field components the excess field produced by the
widened notch leads to a small bending of the actual trap axis
away from the $z$-axis by a distance on the order of $100$~nm.

\begin{figure*}[htbp]
\parbox{5cm}{\caption{(a)~Axial component of the rf field
  $E_{\mathrm{rf, z}}$ on the trap axis as a function of the
  axial coordinate for different deviations $\Delta w$ in the width of one notch in
  an rf carrying electrode. All other notches are kept at $w_n=0.1$~mm. The notches are centered at $z=0$.
  (b)~Maximum amplitude of $E_{\mathrm{rf, z}}$ along the trap axis as a function of $\Delta w$.
  Data points are fitted to show their linear dependence in this range.}\label{notchwidthtol}} \parbox{12.5cm}{
 \begin{flushright}
 \subfloat{\includegraphics[width=0.4387\textwidth]{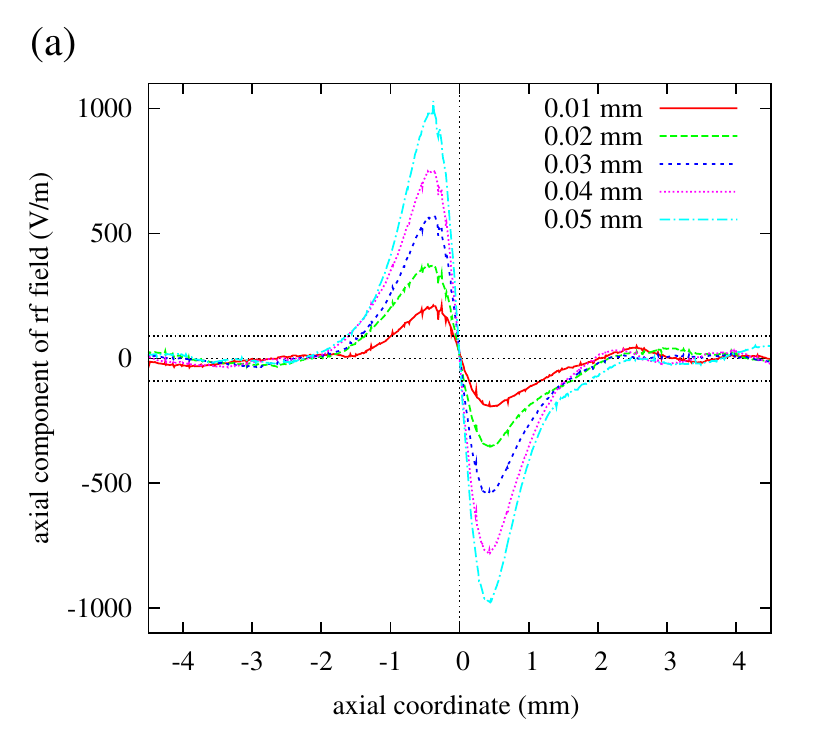}}
 \subfloat{\includegraphics[width=0.2362\textwidth]{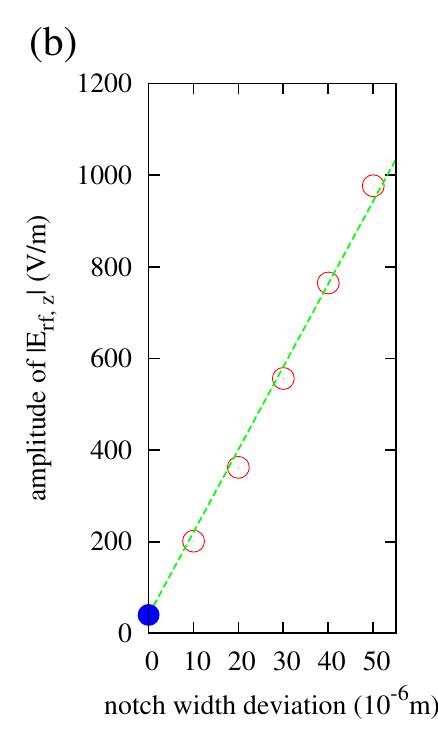}}
 \end{flushright}}
\end{figure*}

Figure~\ref{notchwidthtol}(a) shows the axial rf field component
as a function of the axial coordinate obtained for different
positive deviations $\Delta w$ in the width of one notch from the
width of the notches in the other three electrodes, which are kept
at $w_n=0.1$~mm.
The dependence of the amplitude of the dispersion-shaped feature
in the axial rf field component on the deviation in the width of
one notch is shown in figure~\ref{notchwidthtol}(b).  The line in
the graph resulting from a least square fit suggests, that the
dependence can be regarded to be linear in this range. Assuming a
segment length of $l_s=1$~mm, we deduce a maximum tolerance
 for the notch width in the rf carrying electrode
of $\Delta w=3$~$\mu$m in order to have $|E_{\mathrm{rf, z}}|\leq
90$~V/m in the middle of the segment.  For a segment length of
2~mm the middle of the trap segment is further away from the
segment border and the region where $E_{\mathrm{rf,z}}$ peaks,
such that notches can be cut to less stringent tolerances of
$\Delta w=11$~$\mu$m.

By widening a notch in an rf ground electrode instead of in an rf
carrying electrode we find in our calculations a very similar
effect in the axial component of the rf field on the trap axis,
albeit smaller in amplitude, by roughly 20\% for the here
considered geometry, and of opposite sign.

We also varied the design width $w_n$ of the electrode notches in
the range from 0.1~mm to 0.2~mm and observed a linear increase of
the amplitude of the effect of a widened notch with increasing
$w_n$. The amplitude of the effect was increased 1.8 times for a
design notch width of 0.2~mm as compared to a design notch width
of 0.1~mm.

The calculations in figure~\ref{notchwidthtol} have been carried
out for an electrode thickness of $t_e=0.4$~mm. When we repeated
the calculations for $t_e=0.2$~mm we found only a small increase
in the magnitude of $E_{\mathrm{rf,z}}$  by a factor of 1.07 for a
design notch width of $w_n=0.1$~mm and by a factor of 1.15 for
$w_n=0.2$~mm.

\begin{figure*}[htbp]
\parbox{5cm}{\caption{(a)~Axial component of the rf field
  $E_{\mathrm{rf, z}}$ on
  the trap axis as a function of the axial coordinate for
  different deviations $\Delta z$ in the position of one notch.  In the geometry
  used for the calculations $w_n=0.1$~mm, except for the
  curve mentioned in the legend, where $w_n=0.2$~mm.  (b)~Maximum of $\left|E_{\mathrm{rf, z}}\right|$ as a function
  of $\Delta z$.} \label{notchpostol}} \parbox{12.5cm}{\begin{flushright}
\subfloat{\includegraphics[width=0.4387\textwidth]{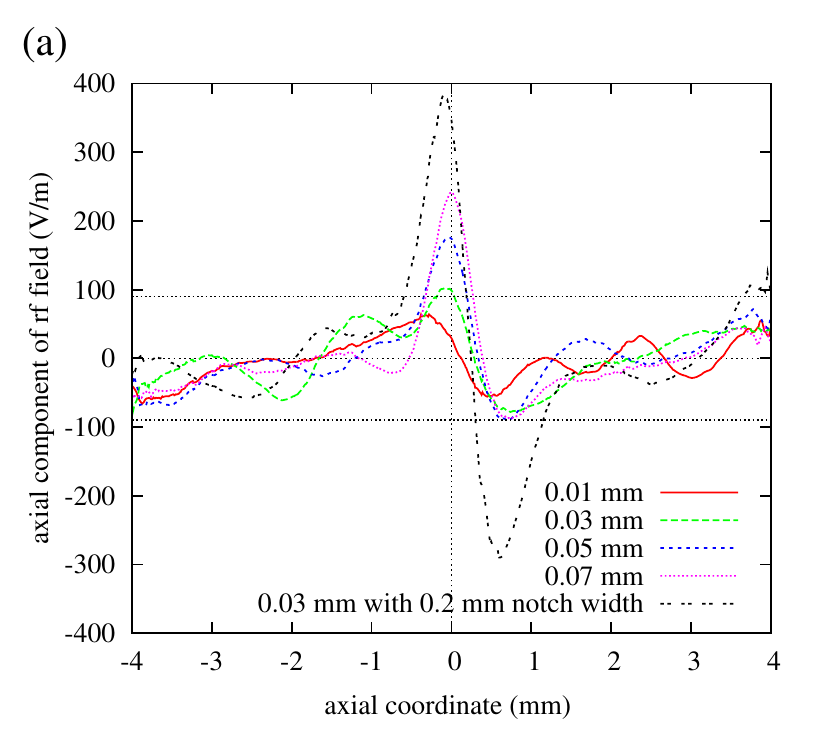}}
\subfloat{\includegraphics[width=0.2362\textwidth]{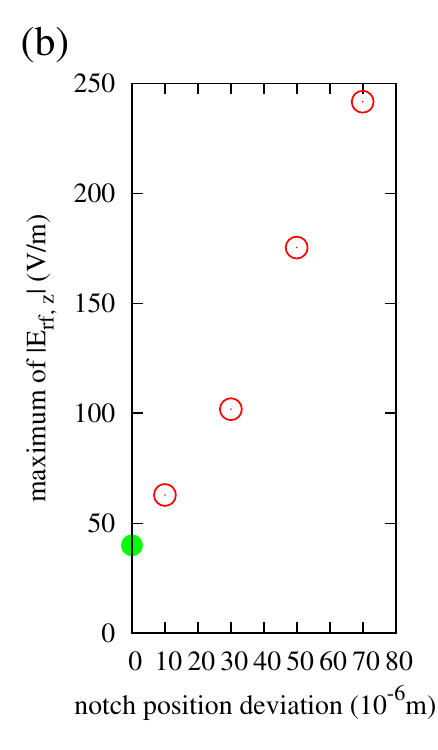}}
\end{flushright}}
\end{figure*}

Second, we consider an inaccuracy $\Delta z$ in the position of
one notch in an otherwise perfectly symmetric electrode assembly.
Figure~\ref{notchpostol} shows the effect on $E_{\mathrm{rf,z}}$
when one notch is displaced in positive $z$ direction.
On the axial component of the rf field the effect is asymmetric
with a maximum located around the center of the notches followed
by a minimum of smaller magnitude located around $z=0.7$~mm.  The
magnitude of the additional axial rf field produced by a position
inaccuracy is considerably smaller than that found for
inaccuracies in the width of a notch. Since the maximum of
$|E_{\mathrm{rf, z}}|$ is also localized on the border of the
segment the trap region in the center of the segment is even less
affected.

For the calculations shown in figure~\ref{notchpostol}(a) the notch
width $w_n=0.1$~mm has been used with exception of one curve,
for which the chosen notch width of 0.2~mm is pointed out in the
legend of the graph. We observed a rapid increase of the maximum of
$|E_{\mathrm{rf,z}}|$ as a function of the notch width, the
displacement of one notch with respect to the others being kept
constant. We want to point out that the residual axial rf field
amplitude occurring in the perfectly symmetric case also grows
rapidly with increasing width of the notches (see
fig.~\ref{notcheffect}) and can contribute significantly here. The
curves in figure~\ref{notchpostol}(a) have been calculated using
an electrode thickness of 0.4~mm. When we used an electrode
thickness of 0.2~mm instead, we did not find a significant
increase in the effect.

Targeting a positioning accuracy $\Delta z$ of better than 10
$\mu$m when aligning the different electrode layers on top of each
other, the axial rf field contribution due to position
inaccuracies will not become critical for a notch width of up to
0.2~mm.
\subsection{Translational misalignment of quadrupole electrodes}\label{toltrans}
We looked into the misalignment of the trap assembly in which one
of the electrode wafers is translated with respect to the other
parallel to the trap axis.  The translation destroys the symmetry
of the geometry at the end of the trap segments, where the notches
in the electrodes of one wafer become displaced with respect to
the corresponding notches in the other wafer. Considering such a
pair of notches of which each is in another wafer, we expect that
the displacement results in an extra axial component of the rf
field halfway between the two notches.  In our FEM calculations
the extra axial rf field contribution was evident between the
wafers in the vicinity of the notches.

The calculations also show that the extra axial field decreases in
magnitude towards the trap axis to finally vanish right on the
axis. Since for the pair of notches situated on the opposite side
of the trap axis the configuration of rf and rf ground electrodes
is reversed on the two wafers, the additional axial rf field
component has opposite sign on both sides of the trap axis. As a
result the effects of the translation from two opposite pairs of
notches cancel each other on the trap axis.  From our calculation
using $w_n=0.1$~mm and $t_e=0.4$~mm we conclude that even at a
distance of 0.05~mm from the trap axis and at $z$ coordinates
corresponding to the location of electrode notches, an axial
translation of 0.05~mm of the wafers with respect to each other
results in a contribution to the axial component of the rf field
smaller than 90~V/m. In the middle of the segment where the ions
are trapped the effect is even smaller.
\subsection{Introduction of extra electrode layers for dc field compensation}\label{xtrcomps}
In order to compensate for dc stray fields in radial
direction, which would displace trapped ions from the null field
line of the rf field, the trap design has to offer a way to apply
independent dc compensation fields in the radial plane in each
trap segment. Two degrees of freedom are required to compensate
for any electric field in the radial plane. Using a small voltage
difference between the rf ground electrodes of the trap segment
generates a dc electric field oriented in the (x,y) plane nearly
along the (1,1) direction. In order to generate a second dc
electric field with a strong component perpendicular to this axis,
we considered two alternative designs.

\begin{figure}[htbp]
\begin{center}
\includegraphics[width=0.35\textwidth]{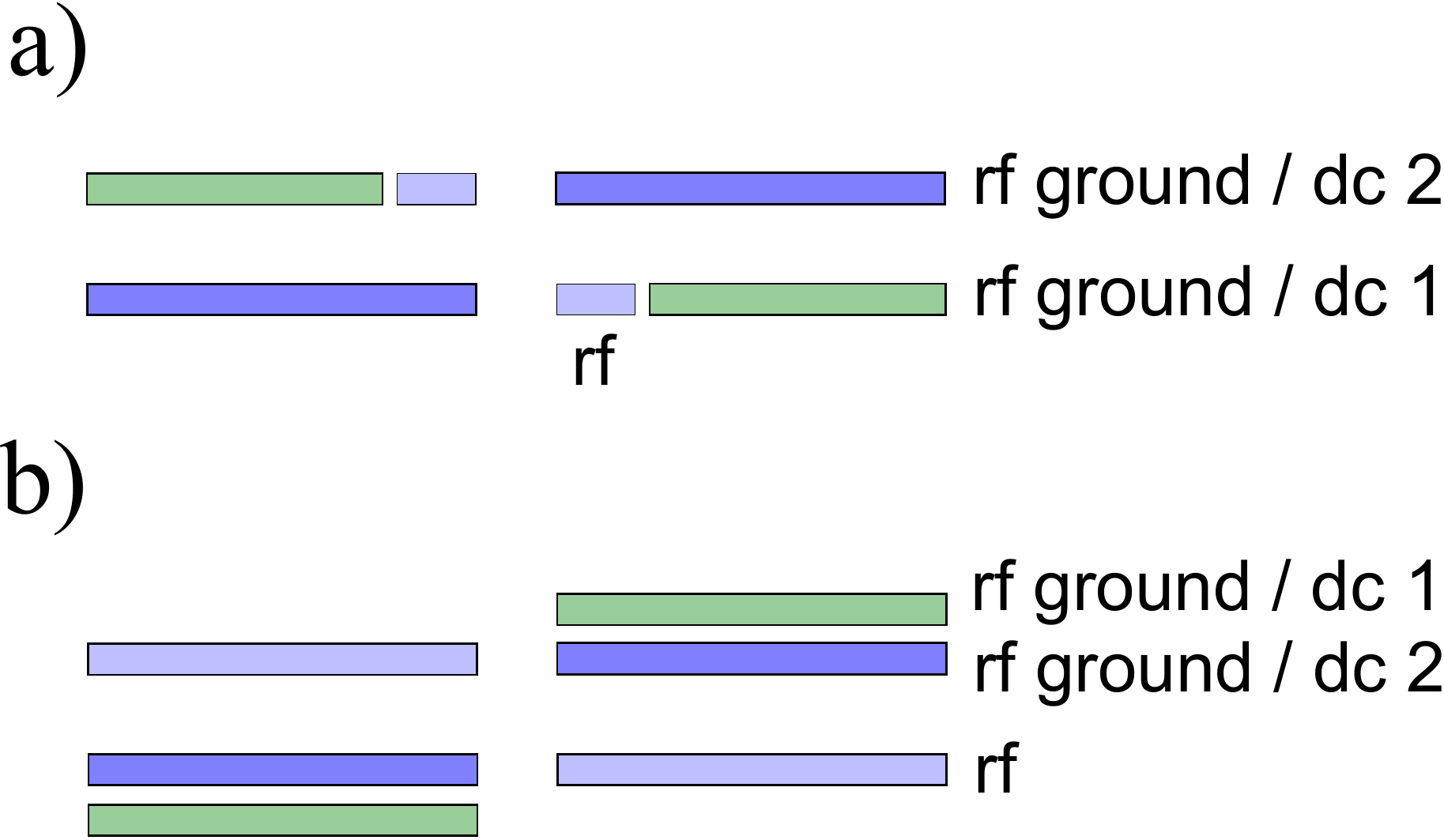}
\caption{Alternative electrode designs to obtain two degrees of
freedom for dc field compensation in the radial ($x,y$) plane.}
\label{xtrcompsdesign}
\end{center}
\end{figure}

In a first design, schematically drawn in
figure~\ref{xtrcompsdesign}(a), we added segmented compensation
electrodes behind a 1~mm wide rf electrode with an insulation gap
of 0.2~mm between rf and compensation electrodes. This design is
compact but bears some strong shielding of the compensation
electrodes by the trap electrodes. The FEM calculation of the dc
field generated in this geometry with an electrode thickness of
0.4~mm, a segment length of 1~mm and a notch width of 0.1~mm gives
a field of 8~V/m in the trap region of the segment using +1~V and
-1~V on the extra compensation electrodes of the segment. Also in
the next and second to next neighboring segments considerably
large dc fields of 4~V/m and 1.4~V/m are found.

A second design consists in expanding the two-layer quadrupole
trap assembly by two additional layers with dedicated segmented
compensation electrodes. Fig.~\ref{xtrcompsdesign}(b) gives a
schematic view of this design. The additional layers are similar
to the other two trap layers only that the rf electrode side is
missing. Each extra layer is placed with a spacing of 0.25~mm on
top and precisely aligned with a segmented rf ground electrode.
The geometry of the extra compensation layers is kept identical to
the inner trap electrodes regarding notch width and electrode
thickness. The FEM calculation of the dc field generated by this
geometry using an electrode thickness of 0.4~mm, a segment length
of 1~mm, and a notch width of 0.1~mm gives a stronger field value
in the trap region of 60~V/m for the voltages of +1~V and -1~V on
the two extra compensation electrodes. The direction of the
generated field vector points mainly  along the $x$-axis with a
component of 3.7~V/m along the $y$-axis. The corresponding field
values obtained in the next and second to next neighboring
segments are 19~V/m and 2.5~V/m, respectively.

The second design has a clear advantage over the first design
regarding the larger field. The field is also well localized on
the trap segment with only a 4\% contribution to the dc field in
the neighboring traps. In addition, considering the relatively
large rf voltage amplitude we envisage to apply, the second design
is also less prone to rf breakdown and surface flashover than the
first. The slightly
larger layer stack reduces the solid angle for optical detection to an
acceptable level.

\begin{figure}[htbp]
\begin{center}
\includegraphics[width=0.5\textwidth]{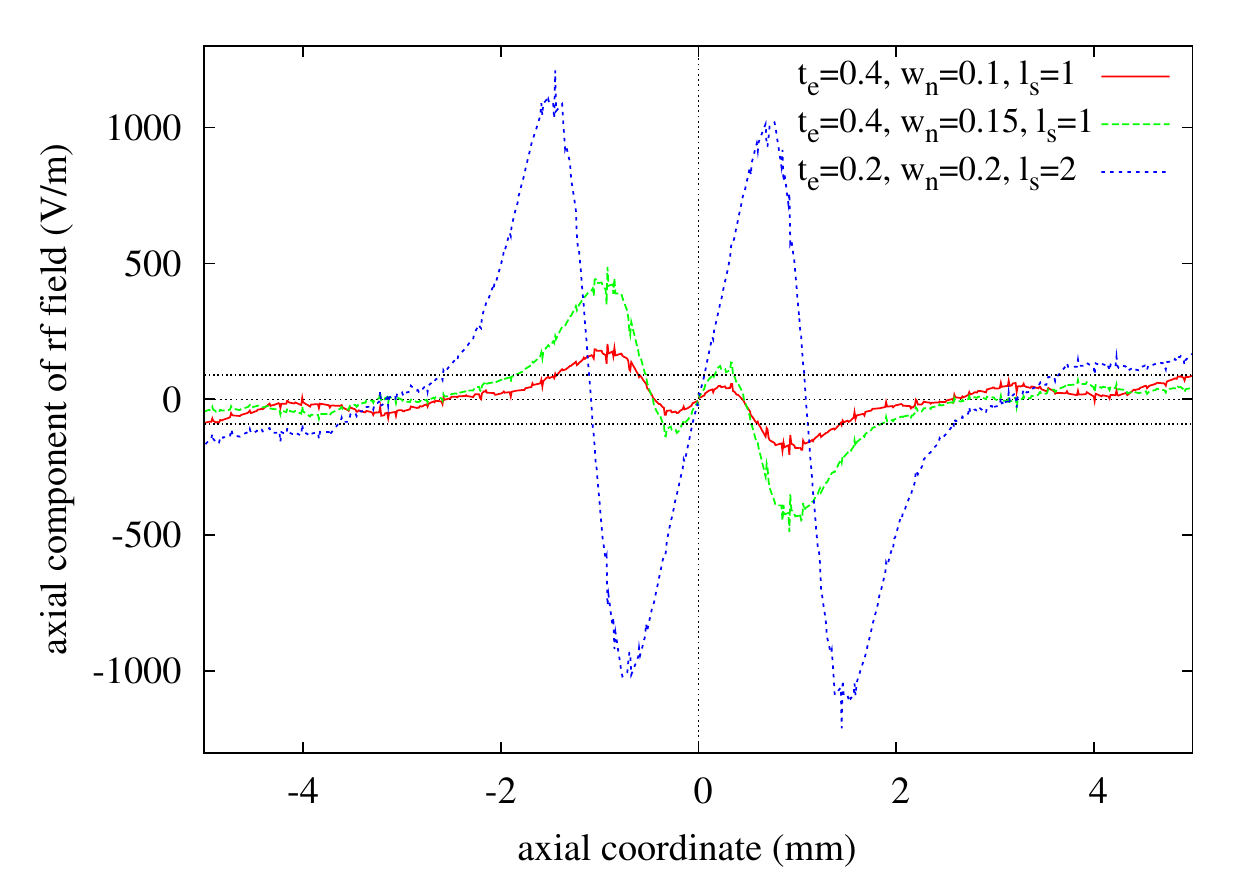}
\caption{(a)~Axial component of the rf field $E_{\mathrm{rf, z}}$
on
  the trap axis  as a function of the axial coordinate $z$ for a
  3-segment trap design with extra compensation electrode
  layers. The center of the middle trap segment is located at
  $z=0$. The curves are calculations for different
  geometrical design parameters: electrode thickness $t_{\mathrm{e}}$,
  notch width $w_{\mathrm{n}}$ and segment length $l_{\mathrm{s}}$
  given in mm in the legend.}
\label{threel_2sl_Ez}
\end{center}
\end{figure}

Since we opt for the second design we investigate the effect of
the extra compensation electrode layers on the rf field. It is
important to examine how the additional compensation electrodes
affect the axial component of the rf field produced by the
electrode notches.  Considering the effect on a single segment
border we calculate the rf field for an electrode geometry with
only two long segments, thus one central discontinuity in the
electrode blades. Adding the extra compensation electrode layers
to the trap geometry clearly increases the amplitude of the axial
rf field component caused by the notches. Compared to calculations
for the same geometry without extra compensation electrodes as
described in section~\ref{slits} (figure~\ref{notcheffect}), we
find that the amplitude of $E_{\mathrm{rf,z}}$ is increased by a
factor of 3.6 for an electrode thickness of 0.2~mm and a notch
width of 0.2~mm, and by a factor of 4.6 for an electrode thickness
of 0.4~mm and notch widths of 0.1~mm as well as 0.15~mm.  Although
the extra compensation layers lead to a considerable increase in
the axial rf field component on the segment borders, the axial rf
field component can still be small enough in the region in the
middle of a trap segment, because here again the effects stemming from
neighboring segment borders partially cancel one another out. An example
for this is given in figure~\ref{threel_2sl_Ez} for
three different design parameters: electrode thickness, notch
width and segment length. Here, a three segment trap was simulated
with the middle segment of length $l_s$ surrounded by two long
outer segments.
In the calculations for all three parameter sets shown in
figure~\ref{threel_2sl_Ez} an axial region of at least
0.12~mm length with $|E_{\mathrm{rf,z}}|<90$~V/m can be found,
which is suitable for clock spectroscopy.

\section{Conclusion and outlook}
\label{sec:5}

We presented an electrode design of a segmented linear Paul trap
optimized for optical clock application and frequency
standards based on the spectroscopy of Coulomb crystals. Using
segmented electrodes the design supports the implementation of
linear arrays of traps, where in each trap segment electrodes are
provided for the compensation of dc electric fields in all three
directions of space. The trap assembly still offers a good optical
access for laser beams and fluorescence detection.

In this design geometry we propose to trap arrays of
$^{115}$In$^+$ ions sympathetically cooled by $^{172}$Yb$^+$. We
showed that with such an optical frequency standard of improved
short-term stability a fractional long-term stability of
$10^{-18}$ can be reached with systematic frequency shifts at the
mHz level. Integration times for the frequency measurement can be
shortened by more than an order of magnitude and requirements for
clock laser stability relaxed significantly.

Of great importance for high precision spectroscopy of a larger
sample of ions is the minimization of excess micromotion of the
ions, which can contribute considerably to the second-order
Doppler shift of the clock transition. In our design we minimized
the critical axial component of the rf field along the trap axis
and studied tolerances for the manufacturing process of the trap.
We identified and quantified the three most critical causes for
excessive magnitudes of the axial rf field component: the finite
length of the quadrupole electrodes; the tolerance on the width of
the notches separating electrode segments; and the tolerance on
the alignment angle of opposing electrode wafers (torsion around
the $x$ axis in our coordinate system). For our choice of geometry
with an 1~mm~$\times$~1~mm spacing between the quadrupole
electrodes, and requiring the axial component of the rf field on
the trap axis to not limit the targeted fractional uncertainty of
the frequency standard of $10^{-18}$,
we can summarize our results on these critical effects as follows.
Using a trap segment length of 1~mm and electrode thickness of
0.2~mm a linear array of six usable traps with at least 10 ions
each is obtained with a total electrode length of 3~cm, whereas
with a total electrode length of 2~cm our design only supports two
usable traps. Choosing the width of the notches that are
separating trap segments to be 0.1~mm and the segment length to be
1~mm, the tolerance on the width of a notch in the electrodes is
$3$~$\mu$m. The tolerance on the angular misalignment of the two
quadrupole electrode layers around the direction orthogonal to the
wafers is less than 0.14~mrad.

The effect of other tolerances in the trap construction, like the
position deviation of a single notch in the electrodes, the
translational displacement along the trap axis of one electrode
layer with respect to the other, and the angular misalignment of
the electrode layers around the $y$ axis, have also been
investigated and quantified. We showed that their effect on the
axial component of the rf field on the trap axis is either
negligibly small or can be easily avoided by applying
state-of-the-art tolerances in the manufacturing process. Further,
we point out that depending on the choice of segment length
boundary effects of the segmentations can cancel each other out,
reducing the on-axis micromotion.

Based on our FEM calculations a prototype trap with two trapping
segments made of Rogers4350B$^{\mathrm{TM}}$ has been constructed and is
currently under test. First measurements of micromotion with
$^{172}$Yb$^+$ ions show a promising low value in agreement with
our simulations. For a scalable high precision trap we are
developing a laser cutting and metallization process for AlN
wafers, in which we plan to investigate an array of $6 \times 10$
ions for an optical frequency standard.

\emph{Acknowledgments:} We thank Ch. Roos for contributing ideas
on trap field calculations and E. Peik for fruitful discussions
and careful reading of the manuscript. This work was supported by
the cluster of excellence QUEST.
%


\begin{thebibliography}{}
\bibitem{Neuhauser} W. Neuhauser et al., Phys. Rev. A \textbf{22,} (1980) 1137.
\bibitem{Werth} F.G. Major, V.N. Gheorghe and G. Werth,  \textit{Charged Particle Traps} (Springer, Berlin 2010).
\bibitem{Gabrielse} L. S. Brown and G. Gabrielse, Rev. Mod. Phys. \textbf{58,} 233--311 (1986).

\bibitem{Bergquist_Rosenband_Science} T. Rosenband et al., Science \textbf{319,} (2008) 1808.
\bibitem{Chou} C. W. Chou et al., Phys. Rev. Lett. \textbf{104,} (2010) 070802.

\bibitem{Itano} W. Itano et al., Phys. Rev. A \textbf{47,} (1993) 3554.

\bibitem{Meiser} D. Meiser et al, Phys. Rev. Lett. \textbf{102,} (2009) 163601.
\bibitem{Millo} J. Millo et al., Phys. Rev. A \textbf{79,}  (2009) 053829.
\bibitem{Legero} T. Legero et al., JOSA B \textbf{27,} (2010) 914.

\bibitem{Bjerhammer} A. Bjerhammar, Bull. Geodesique \textbf{59,} (1985) 207.
\bibitem{Pavlis} N. K. Pavlis and M. A. Weiss, \textit{Frequency Control Symposium,
2007} IEEE International (2007) 642.


\bibitem{Riehle} F. Riehle, \textit{Frequency Standards: Basics and Applications} (Wiley-VCH, Weinheim 2005).

\bibitem{Prestage} J. D. Prestage et al., \textit{Frequency Control Symposium,
2007} IEEE International (2007) 1112.
\bibitem{Champenois} C. Champenois et al., Phys. Rev. A \textbf{81,} (2010) 043410.

\bibitem{Schmidt} P. Schmidt et al., Science \textbf{309,}  (2005) 749.
\bibitem{Rosenband2007} T. Rosenband et al., Phys. Rev. Lett. \textbf{98,} (2007) 220801.

\bibitem{BlattWineland} D. Wineland and R. Blatt, Nature \textbf{453,} (2008) 1008.
\bibitem{Blatt} Th. Monz et al., arXiv:1009.6126v1 (2010).
\bibitem{Monroe1995} Ch. Monroe et al., Phys. Rev. Lett \textbf{75,} (1995), 4011.
\bibitem{Amini} Amini et al., arXiv:0812.3907v1 (2008).
\bibitem {Kielpinski} D. Kielpinski et al., Nature \textbf{417,}  (2002) 709.

\bibitem{Turchette} Q. A. Turchette et al., Phys. Rev. A \textbf{61,} (2000) 063418.
\bibitem{Berkeland} D. Berkeland et al., J. Appl. Phys. \textbf{83,} (1998) 5025.

\bibitem{Dicke} R. H. Dicke, Phys. Rev. \textbf{89,} (1953) 472.

\bibitem{tbp} K. Pyka et al., in preparation for New J. Phys.

\bibitem{Zipkes} Ch. Zipkes et al., Nature \textbf{464,} (2010) 388.

\bibitem{Peik} E. Peik et al., Phys. Rev. A {\bf 49}, (1994) 402.
\bibitem{Liu} T. Liu, Y. Wang, V. Elman, A. Stejskal, Y. Zhao, J. Zhang, Z. Lu, L. Wang, R. Dumke, T. Becker and H. Walther, \textit{Frequency Control Symposium, 2007} IEEE International (2007) 407.
\bibitem{Fortson} J. A. Sherman, W. Trimble, S. Metz, W. Nagourney and N. Fortson, arXiv:physics/0504013v2 (2005).


\bibitem{Schwedes} Ch. Schwedes et al., Appl. Phys. B {\bf 76},  (2003) 143.


\bibitem{Sugiyama} Sugiyama et al., Phys. Rev. A \textbf{55,} (1997) R10.

\bibitem{Safronova} M.S. Safronova and M.G. Kozlov, arXiv:1105.3233 (2011).
\bibitem{Rosenband_BBR} T. Rosenband et al., Proceedings of the 20th European Time and Frequency Forum (2006) 289.

\bibitem{Peik_1999} E. Peik et al., Phys. Rev. A \textbf{60,} (1999) 439.
\bibitem{Deslauriers} L. Deslauriers et al., Phys. Rev. Lett. \textbf{97,} 103007 (2006).
\bibitem{Epstein} R. J. Epstein et al., Phys. Rev. A \textbf{76,} (2007) 033411.

\bibitem{Madej} J. E. Bernard et al., Opt. Comm. \textbf{150,} (1998) 170.

\bibitem{Becker} Th. Becker et al., Phys. Rev. A \textbf{63,} (2001) 051802.

\bibitem{Lin} Lin et al., Europ. Phys. Lett. \textbf{86,} (2009) 60004.

\bibitem{Naegerl} H. C. N\"{a}gerl et al., \textit{The Physics of Quantum Information} (Springer, Berlin 2000).
\bibitem{Schulz} S. Schulz et al., New J. Phys. \textbf{10,} (2008) 045007.
\bibitem{Madsen} M. J. Madsen et al., Appl. Phys. B \textbf{78,} (2004) 639.

\bibitem{Wineland_Monroe} D. J. Wineland et al., J. Res. Nat. Bur. Stand. Tech. \textbf{103,} (1998) 259.
\bibitem{Steane} A. Steane, Appl. Phys. B \textbf{64,}  (1997) 623.

\bibitem{Dubin} D. H. E. Dubin, Phys. Rev. Lett. \textbf{71,}  (1993) 2753.

\bibitem{Totsuji} H. Totsuji and J.-L. Barrat, Phys. Rev. Lett. \textbf{60,}  (1988) 2484.

\bibitem{Schiffer} J. P. Schiffer, Phys. Rev. Lett. \textbf{70,} (1993) 818.

\bibitem{Wineland_Itano_1987} D. J. Wineland et al., Phys. Rev. A \textbf{36,} (1987) 2220.

\bibitem{Goering_fem} I. Babuska and T. Strouboulis, \textit{The finite element method and its reliability}, (Oxford University Press, New York 2001).


\end{thebibliography}
\end{document}